\documentclass[10pt,prd,aps,eqsecnum,twocolumn,showpacs,nofootinbib,superscriptaddress,preprintnumbers,floatfix,]{revtex4-1}

\usepackage[raggedright]{subfigure}
\usepackage{amsmath}
\usepackage{amsfonts}
\usepackage{graphicx}
\usepackage[usenames]{color}

\newcommand{\be}{\begin{equation}}
\newcommand{\ee}{\end{equation}}
\newcommand{\bea}{\begin{eqnarray}}
\newcommand{\eea}{\end{eqnarray}}

\newcommand{\tr}{\,{\rm tr}}

\newcommand{\nn}{\nonumber}

\begin{document}

\title{Instabilities of an anisotropically expanding non-Abelian plasma:\\ 
         3D+3V discretized hard-loop simulations}

\preprint{TUW-12-16}

\author{Maximilian Attems}  
\author{Anton Rebhan}
\affiliation{Institut f\"ur Theoretische Physik, Technische Universit\"at Wien,
Wiedner Hauptstrasse 8-10, A-1040 Vienna, Austria}
\author{Michael Strickland}
\affiliation{Physics Department, Gettysburg College, Gettysburg, PA 17325 USA}
\affiliation{Frankfurt Institute for Advanced Studies, Ruth-Moufang-Strasse 1,
D-60438, Frankfurt am Main, Germany}

\date{\today}

\begin{abstract}
We study the (3+1)-dimensional 
evolution of non-Abelian plasma instabilities in the
presence of a longitudinally expanding background of hard
particles using the discretized hard loop framework.  The free
streaming background dynamically generates a momentum-space
anisotropic distribution which is unstable to the rapid growth of
chromomagnetic and chromoelectric fields.  These fields produce
longitudinal pressure that works to isotropize the system.
Extrapolating our results to energies probed in ultrarelativistic heavy-ion 
collisions we find, however, that a pressure anisotropy persists for a few fm/c.
In addition, on time scales relevant to heavy-ion collisions
we observe continued growth of plasma instabilities 
in the strongly non-Abelian regime.  Finally, we find that the longitudinal
energy spectrum is well-described by a Boltzmann distribution with 
increasing temperature at intermediate time scales. 
\end{abstract}

\maketitle

\newpage 

\section{Introduction}
\label{sec:intro}

One of the major outstanding questions in the theoretical understanding of ultrarelativistic heavy ion collisions 
concerns the thermalization and isotropization of the quark gluon plasma.  Empirical evidence in favor of fast 
thermalization and isotropization was provided by ideal relativistic hydrodynamical models.  The success of these models 
to describe the collective flow observed at the Relativistic Heavy Ion Collider (RHIC) suggested that one generated thermal 
and isotropic matter at time scales on the order of 0.5 fm/c after the initial nuclear impact 
\cite{Huovinen:2001cy,Hirano:2002ds,Tannenbaum:2006ch, Kolb:2003dz}. 
Based on this success there was a concerted effort to include corrections due to the finite
shear viscosity of the plasma \cite{Muronga:2001zk,Muronga:2003ta,Muronga:2004sf,Baier:2006um,%
Romatschke:2007mq,Baier:2007ix,Dusling:2007gi,Luzum:2008cw,Song:2008hj,Heinz:2009xj,%
El:2009vj,PeraltaRamos:2009kg,PeraltaRamos:2010je,Denicol:2010tr,Denicol:2010xn,%
Schenke:2010rr,Schenke:2011tv,Shen:2011eg,Bozek:2011wa,Niemi:2011ix,Niemi:2012ry,Bozek:2012qs,%
Denicol:2012cn}.  Second order viscous hydrodynamics is now widely used to model collisions at
both RHIC and the Large Hadron Collider (LHC).

In recent years, however, studies have shown that there is 
an insensitivity to the assumed momentum space anisotropy of the plasma, with the data also being consistent 
with initially large momentum-space anisotropies \cite{Shen:2011eg,Ryblewski:2012rr}.  
In addition, studies based on the conjectured anti de Sitter/conformal field theory (AdS/CFT) correspondence 
have shown that, although viscous hydrodynamical behavior emerges quickly in the strong coupling limit,
there are still sizable momentum-space anisotropies present that persist over the entire lifetime of
the plasma \cite{Heller:2011ju,Heller:2012je,Wu:2011yd}.  
Based on this, extensions of viscous hydrodynamics that can accommodate large
momentum-space anisotropies have been developed \cite{Martinez:2009mf,Martinez:2009mf,%
Florkowski:2010cf,Martinez:2010sc,Ryblewski:2010bs,Martinez:2010sd,Ryblewski:2011aq,%
Florkowski:2011jg,Martinez:2012tu,Ryblewski:2012rr}.  Currently the question of the degree of momentum-space 
isotropy of the quark gluon plasma generated in heavy ion collisions is an open question.  In this paper we study the 
role played by collective unstable modes of the chromomagnetic and chromoelectric fields in restoring momentum-space 
isotropy of an expanding quark gluon plasma (QGP).

It has been shown using both kinetic theory and diagrammatic methods that when the local particle distribution 
function of a weakly-coupled QGP is anisotropic in momentum space, the system is unstable 
to the rapid growth of soft gauge fields \cite{Heinz:1985vf,Mrowczynski:1988dz,Pokrovsky:1988bm,%
Mrowczynski:1993qm,Mrowczynski:2000ed,Randrup:2003cw, Romatschke:2003ms,Arnold:2003rq,%
Arnold:2004ih,Romatschke:2004jh,Arnold:2004ti}.  This instability has been dubbed the chromo-Weibel
instability in reference to the Abelian analogue of this instability first discussed by Weibel \cite{Weibel:1959}.
In the weak-field regime the chromo-Weibel instability initially causes exponential growth of transverse
chromomagnetic and chromoelectric fields; however, due to non-Abelian interaction between the fields, 
exponentially growing longitudinal chromomagnetic and
chromoelectric fields are induced that grow at twice the rate of the transverse field configurations.
As a result, one finds strong gauge field self-interaction at late times due to high-amplitude chromoelectric
and chromomagnetic fields and in order to reach quantitative conclusions numerical simulations are necessary.

The initial numerical studies of the time evolution of the
chromo-Weibel instability were performed assuming a static
momentum-space anistropic (non-expanding) system and utilized
discretizations of the gauge-invariant hard-loop action.  The
hard-loop action used includes the self-consistent gauge-invariant
modification of all $n$-point functions in the hard-loop limit
\cite{Mrowczynski:2004kv}.  The resulting discretized dynamical
equations were solved in temporal axial gauge using a regular lattice
to describe space and either a discrete lattice
\cite{Rebhan:2004ur,Rebhan:2005re,Ipp:2010uy} or an expansion in
spherical harmonics \cite{Arnold:2005vb,Arnold:2005qs,Arnold:2005ef,%
  Bodeker:2007fw,Arnold:2007cg} to describe the velocity space of the
hard particles.  From the three-dimensional static box simulations one
found that exponential field growth ceased when the vector potential
amplitude became on the order of $A_{\rm non-Abelian} \sim p_{\rm s}/g
\sim \sqrt{f_h} p_h$, where $p_h$ is the characteristic momentum of
the hard particles, e.g. $p_h \sim Q_s$ for color glass condensate
(CGC) initial conditions, $f_h$ is the angle-averaged occupancy at the
hard scale, and $p_s$ is the characteristic soft momentum of the
fields ($p_s \sim g \sqrt{f_h} p_h$).  This partial-saturation occurs
at a scale where the chromo-fields are not yet strong enough to have
O(1) effects on the hard particle distribution, suggesting that
isotropization in non-Abelian plasmas is parametrically slower than in
the Abelian case.  After the exponential growth ceased, a slower
linear growth of field energy densities was observed.  This linear
growth was associated with a cascade of energy pumped into the soft
modes to higher momentum modes through nonlinear gauge-field
self-interactions \cite{Arnold:2005vb,%
  Arnold:2005ef,Arnold:2005qs}.  The resulting spectrum of soft gauge
field excitations was shown to have a power law spectrum scaling like
$f \sim \alpha_s^{-1} p_s^{-2}$ for SU($N_c$) with $N_c \in
\{2,3,4,5\}$ \cite{Arnold:2005qs,Strickland:2007fm,Ipp:2010uy}.
{
  Studies using classical-statistical Yang-Mills simulations also
  found saturation of gauge-field growth with an associated
  gauge-field power-law spectrum; however, these studies found
  saturation only in a regime where back-reaction on the hard modes is
  already strong, with a different scaling consistent with $f \sim
  \alpha_s^{-1} p_s^{-4/3}$
  \cite{Berges:2007re,Berges:2008mr,Berges:2008zt,Berges:2009bx,Berges:2012iw}.
}

The presence of instabilities in weakly-coupled momentum-space
anisotropic systems seems to be generic and independent of the
hard-loop approximation, the gauge group, and, in large part, the type
of theory considered (including, of course, the weak-coupling limit of
supersymmetric gauge theories \cite{Czajka:2010zh,Czajka:2012gq}).
They have been observed in numerical solutions to the full
Boltzmann-Vlasov equations that go beyond the hard-loop approximation
\cite{Dumitru:2005gp,Dumitru:2006pz,Dumitru:2007rp}.  {
  As mentioned previously, analogous instabilities have been observed
  in numerical simulations of pure classical-statistical Yang-Mills
  dynamics
  \cite{Berges:2007re,Berges:2008mr,Berges:2008zt,Berges:2009bx,Berges:2012iw}.
  As a result, obtaining a detailed understanding of the chromo-Weibel
  instability's effect on the isotropization and thermalization of the
  matter created in ultrarelativistic heavy ion collisions is of
  upmost importance.}  There have been many works that have addressed
pieces of the puzzle
\cite{Arnold:2003rq,Bodeker:2005nv,Mrowczynski:2006ad,%
  Strickland:2007fm,Mrowczynski:2009gf,Carrington:2010sz}.  Recently
there has been a highly impressive effort to parametrically estimate
the effect of plasma instabilities on the quark gluon plasma
thermalization time \cite{Kurkela:2011ti,Kurkela:2011ub}; however,
being a parametric estimate it does not yet fully answer the question
or lend itself to extrapolations to realistic couplings.

In order to understand the precise role the chromo-Weibel instability plays in ultrarelativistic heavy ion
collisions it is necessary to include the effect of the strong longitudinal expansion of the matter, 
particularly during its earliest stages.  For
the first few fm/c of the quark gluon plasma's lifetime the longitudinal expansion dominates the transverse
expansion which only starts to become important at time scales on the order of 4-5 fm/c.  Therefore, to good
approximation, one can understand the early time dynamics of the quark gluon plasma by only considering
longitudinal dynamics.  The first study to look at the effect of longitudinal expansion was done in the
context of pure Yang-Mills dynamics initialized with CGC initial conditions onto which
small-amplitude rapidity fluctuations were added \cite{Romatschke:2005pm,Romatschke:2006nk}.  
The initial small-amplitude fluctuations result from quantum corrections to the classical dynamics 
\cite{Fukushima:2006ax,Dusling:2011rz}.  Numerical studies have shown that adding spatial-rapidity 
fluctuations results in growth of chromomagnetic and chromoelectric fields with amplitudes
$\sim \exp(2 m_D^0\sqrt{\tau/Q_s})$ where $m_D^0$ is the initial Debye screening mass and $\tau$
is the proper time.  This growth with $\exp(\sqrt{\tau})$ was predicted by Arnold et al. based on 
the fact that longitudinal expansion dilutes the density, thereby causing the chromo-Weibel unstable 
growth rate decrease in time \cite{Arnold:2003rq}.  

Since the pioneering study of Refs.~\cite{Romatschke:2005pm,Romatschke:2006nk} others are
now investigating the evolution of instabilities in classical Yang-Mills \cite{Fukushima:2011nq,Berges:2012iw}
and scalar $\phi^4$ \cite{Dusling:2012ig} including longitudinal expansion.  In addition, a parallel effort to 
incorporate longitudinal expansion into the hard-loop framework was begun with the first results being semi-analytic
solutions for Abelian theories that also showed the characteristic $\exp(\sqrt{\tau})$ growth seen in
the earlier classical Yang-Mills simulations, as well as rather complex early-time 
behavior \cite{Romatschke:2006wg}.   In the hard-loop framework
the longitudinal expansion has thus far been included only in the limit that the hard particles are
free streaming.  In this case it is possible to introduce a set of auxiliary variables 
similar to the static hard-loop $W$ fields which account for the time-evolving momentum-space
anisotropy of the hard particle distribution.  

The Abelian semi-analytic solutions of 
Ref.~\cite{Romatschke:2006wg} were shortly followed by numerical solutions of the resulting coupled 
SU(2) Vlasov-Yang-Mills equations in the simplified case that the vector potential ${\bf A}$ and its 
conjugate momenta ${\bf \Pi}$ were homogeneous in the transverse directions \cite{Rebhan:2008uj}.  
Coupling these transversally-homogeneous fields to the fully 
three-dimensional hard-particle velocity distribution resulted in ``1D+3V'' simulations of the 
resulting dynamics.\footnote{Since, in practice, the ultrarelativistic limit $|{\bf v}| \rightarrow 1$ is used, 
the three-dimensional velocity space is further reduced to a two-dimensional space (the surface of a three-sphere).}
This study found that, in the case of non-Abelian SU(2) fields, one also 
observed growth with $\exp(\sqrt{\tau})$ that was only briefly curtailed when the magnitude
of the transverse and longitudinal gauge field energies became of the same order.  In addition,
the 1D+3V simulations did not see a Kolmogorov cascade at late times.

The problem with such dimensionally-reduced studies is that they can be misleading.  In fact, one finds 
in the static box case very different late time behavior if one allows for either effective one-dimensional
dynamics or fully three-dimensional dynamics.  One is therefore motivated to determine the full 
3D+3V dynamics in the presence of a longitudinally expanding background.
In addition, since the 1D+3V paper was written it was realized that the initial conditions used were
not sufficiently generic and that including initial current fluctuations
dramatically reduces the previously observed delayed onset of growth of unstable modes
\cite{Rebhan:2009ku}.  One would therefore like to also use this type of initial condition in the 
full study.  

In this paper, we present the necessary 3D+3V dynamical equations for so-called
hard-expanding-loops (HEL), discretize them in $\tau$-$\eta$-${\bf x}_\perp$ coordinates, and solve them 
numerically.  For this purpose we use anisotropic lattices with spatial sizes on the order of  
$N_\perp^2 \times N_\eta \sim 40^2 \times 128$.  At each point on the lattice we also have
auxiliary fields ${\cal W}$ that are discretized on a velocity-lattice with size 
$N_\phi \times N_u \sim 32 \times 128 $ amounting to 4096 auxiliary
fields per lattice site.  Needless to say this presents a computational challenge that requires 
parallelization of the resulting code.  For the initial conditions we use variants of the initial conditions
specified in Ref.~\cite{Rebhan:2009ku} in which we have added the possibility of initializing an
adjustable spectrum of discrete longitudinal fluctuations.  As in our previous studies, the dimensional 
parameters necessary to fix the initial conditions such as the gluon number density etc.\ are obtained within 
the CGC framework.

We find that, apart from a delay of the onset of the unstable mode growth due to transverse dynamics,
the overall behavior of the three-dimensional solutions is very similar to the one-dimensional case.  We
find that the chromo-Weibel instability acts to restore isotropy
in the system by inducing large longitudinal field pressure.  In contrast to the fixed-anisotropy
3D+3V studies, we do not see a saturation of the instability on time scales relevant for heavy ion collisions.
In order to address the question of the spectrum of the resulting field
configurations we study the longitudinal Fourier-modes of the energy
density.  We find that the longitudinal energy spectrum looks like a Boltzmann distribution while remaining
anisotropic in momentum space.  Extrapolating to energies appropriate for LHC collisions,
we find that the momentum-space anisotropy persists for approximately 6~fm/c.
We show that the isotropization time is primarily determined by the assumed magnitude of initial 
current fluctuations.

The structure of the paper is as follows:  In Section \ref{sec:generalities} we briefly review the expectations
one has for unstable mode growth in an expanding background.  In Section \ref{sec:eom} we review the
derivation of the hard-loop equations of motion in a longitudinally free streaming expanding background.  In
Section \ref{sec:initialconditions-continuum} we discuss the method we used to fix the physical scales in
our simulation and discuss the initial conditions used. In Section \ref{sec:observables} we define the
various observables that we will measure during the lattice evolution. In Section \ref{sec:results}
we present our main results and interpret our findings.  In Section \ref{sec:conclusions} we conclude 
and give an outlook for the future. In three appendices we collect details concerning the numerical
solution of the lattice equations of motion.

\section{General discussion}
\label{sec:generalities}

\begin{figure}[t]
\includegraphics[width=0.5\textwidth]{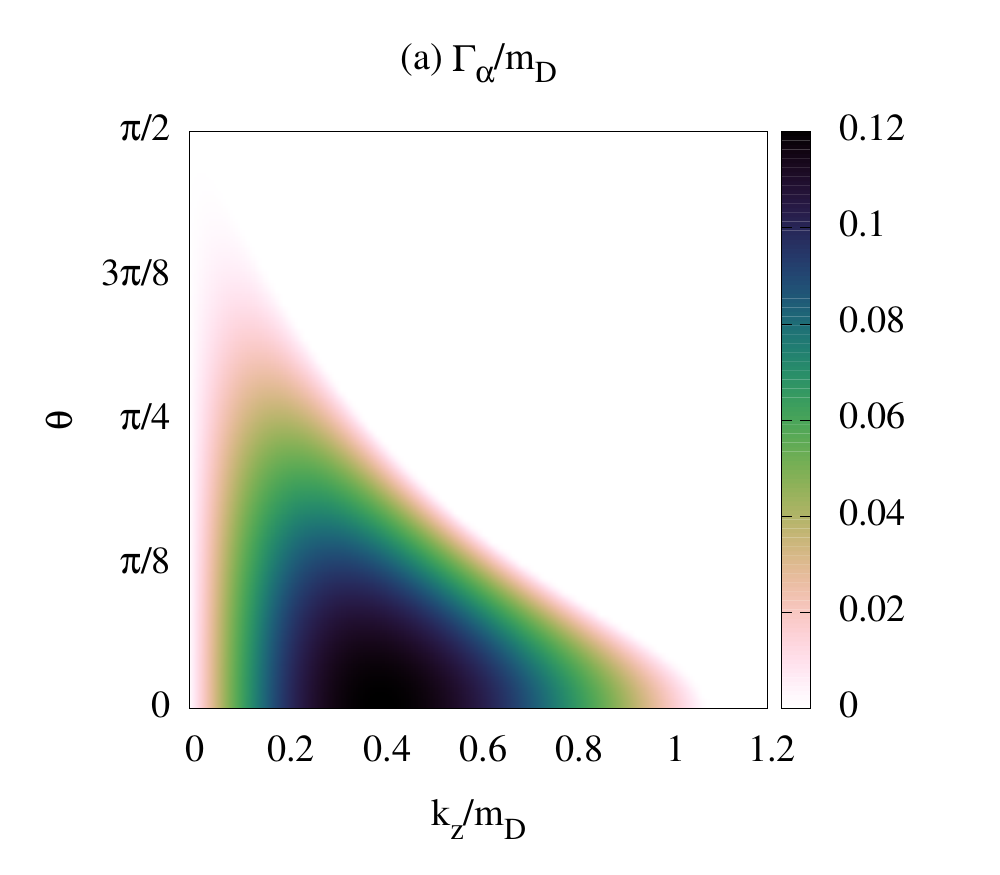}
\includegraphics[width=0.5\textwidth]{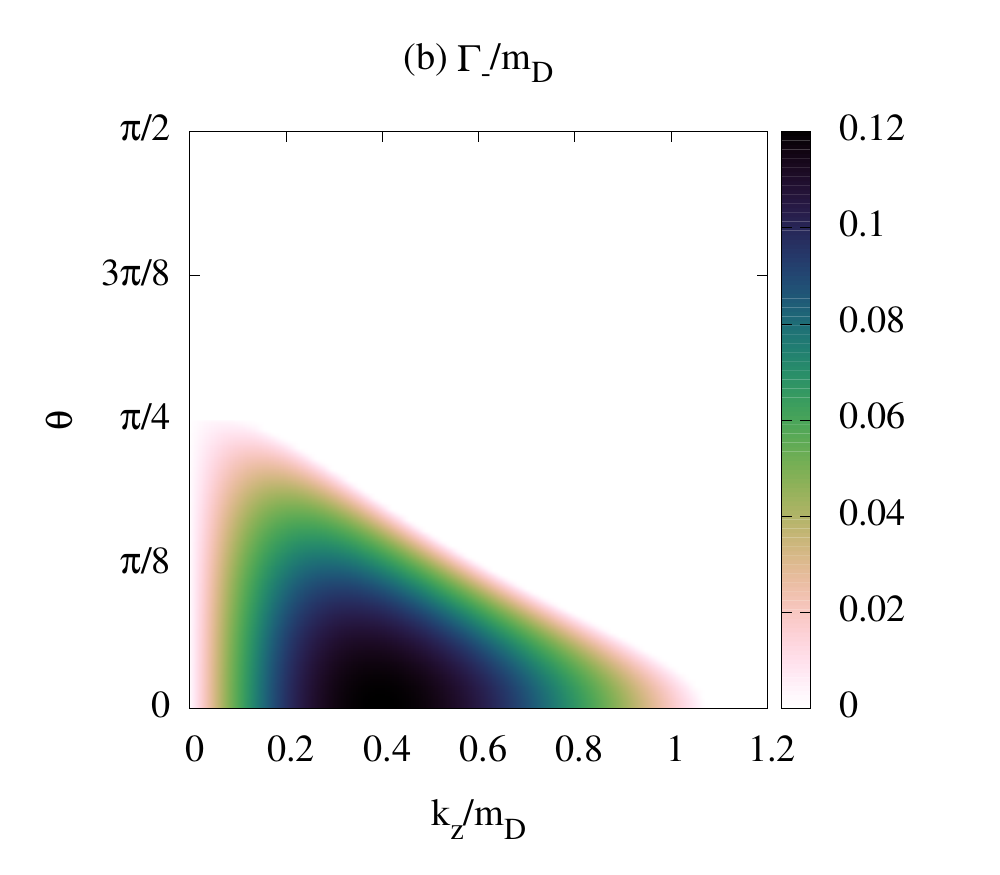}
\caption{(Color online)
Unstable mode growth rates (a) $\Gamma_\alpha/m_D$ and (b) $\Gamma_-/m_D$ for $\xi=10$ as a function of 
$k_z/m_D$ and $\theta = \arctan(k_T/k_z)$ where $m_D$ is the Debye mass at the proper time $\tau_{\rm iso}$. 
}
\label{fig:unstablemodes2d}
\end{figure}

Before proceeding to the presentation of the hard loop equations of motion and their subsequent numerical
solution, we will quickly review the presence of instabilities in a momentum-space anisotropic plasma and
consider how this changes in an expanding plasma.
In a longitudinal free streaming expansion the soft scale is time-dependent.  Since the density of the 
free streaming particles drops like $n \sim 1/\tau$ and $m_D^2(\tau) \propto n/p_{\rm hard}$, we have
\be
m_D(\tau) \sim m_D \left(\frac{\tau}{\tau_{\rm iso}}\right)^{-1/2} \, ,
\ee
where $m_D$ is the ``isotropic'' Debye mass defined at a time $\tau = \tau_{\rm iso}$.

At a given proper time we can quantify the degree of plasma anisotropy via $\xi$
\be
\xi = \frac{1}{2} \frac{\langle p_T^2 \rangle}{\langle p_z^2 \rangle} -1 \, ,
\ee
where $p_T$ and $p_z$ are the transverse and longitudinal (beamline direction) particle momenta in the local reference 
frame.  For a longitudinal free streaming expansion $p_T$ is constant while $p_z \sim 1/\tau$ and as a result one has
$\xi_{\rm f.s.} = (\tau/\tau_{\rm iso})^2 - 1$.~\footnote{The magnitudes of $p_T$ and $p_z$ stated are the 
``expected'' values for the transverse and longitudinal momentum of a particle in the system.  These can be be 
defined formally as $p_T = \sqrt{\langle p_T^2\rangle} \; n_{\rm iso}/n$ and $p_z = \sqrt{\langle p_z^2\rangle} 
\; n_{\rm iso}/n$ where the averages represent integrals using the one-particle distribution and $n$ is the number 
density.}

As we will discuss in Section \ref{sec:initialconditions-continuum}, we assume that a plasma description becomes 
possible after a finite point in proper time $\tau_0$. The ratio $\tau_{\rm iso}/\tau_0$ then parametrizes the initial 
momentum-space anisotropy. If this were equal to one, the plasma would start out isotropic and become anisotropic 
with $\xi>0$ at subsequent times. However, motivated by the results obtained within the CGC framework \cite{Lappi:2006fp} 
we consider the case that the plasma already has a strong oblate  ($\xi >0$)  momentum anisotropy at $\tau_0$, 
which will be modeled by having $\tau_{\rm iso}\ll \tau_0$ regardless of the fact that a plasma description is certainly 
not possible at times earlier than $\tau_0$. By the same token, $m_D$, the isotropic Debye mass at the (fictitious) 
time $\tau_{\rm iso}$, is just a parameter characterizing our free streaming background of hard plasma particles.

At a given proper time $\tau$, and hence fixed plasma anisotropy, there 
is a three-dimensional band of soft unstable modes associated with a fluctuation wave vector ${\bf k}$.  
For an oblate distribution the unstable modes with the largest growth rate have ${\bf k} \parallel \hat{\bf n}$
where $\hat{\bf n}$ is the anisotropy direction \cite{Romatschke:2003ms}.
The oblate unstable modes can be classified as either transverse magnetic ($\alpha$) or mixed (-) modes.
The mixed modes with finite transverse momentum extend out from the anisotropy direction to a fixed angle of 
$\theta = \arctan(k_T/k_z) = \pi/4$ beyond which they are stable.  The $\alpha$-modes, on the other hand, are
unstable for any transverse momentum.  

In Fig.~\ref{fig:unstablemodes2d} we show the range of unstable modes for both types of modes.  We show the 
case of $\xi=10$ with the understanding that the qualitative features are the same for all $\xi > 0$.
In a longitudinally expanding plasma longitudinal 
momenta are redshifted in time, but transverse momenta are unaffected.  As a result, the mixed unstable modes which 
have any finite transverse momentum will eventually become stable.  The $\alpha$-mode growth rate decreases
rapidly as one increases $\theta$, so while they are technically unstable at all times, the growth rate of any mode
which is not purely longitudinal becomes negligible at late times.  Thus, at late times the system will be dominated 
by the dynamics of unstable modes with (nearly) longitudinal wave vectors.\footnote{For a more detailed discussion of the dynamics of stable and 
unstable modes in an anisotropically expanding plasma see Ref.~\cite{Rebhan:2009ku}.}

\begin{figure}
\includegraphics[width=0.5\textwidth]{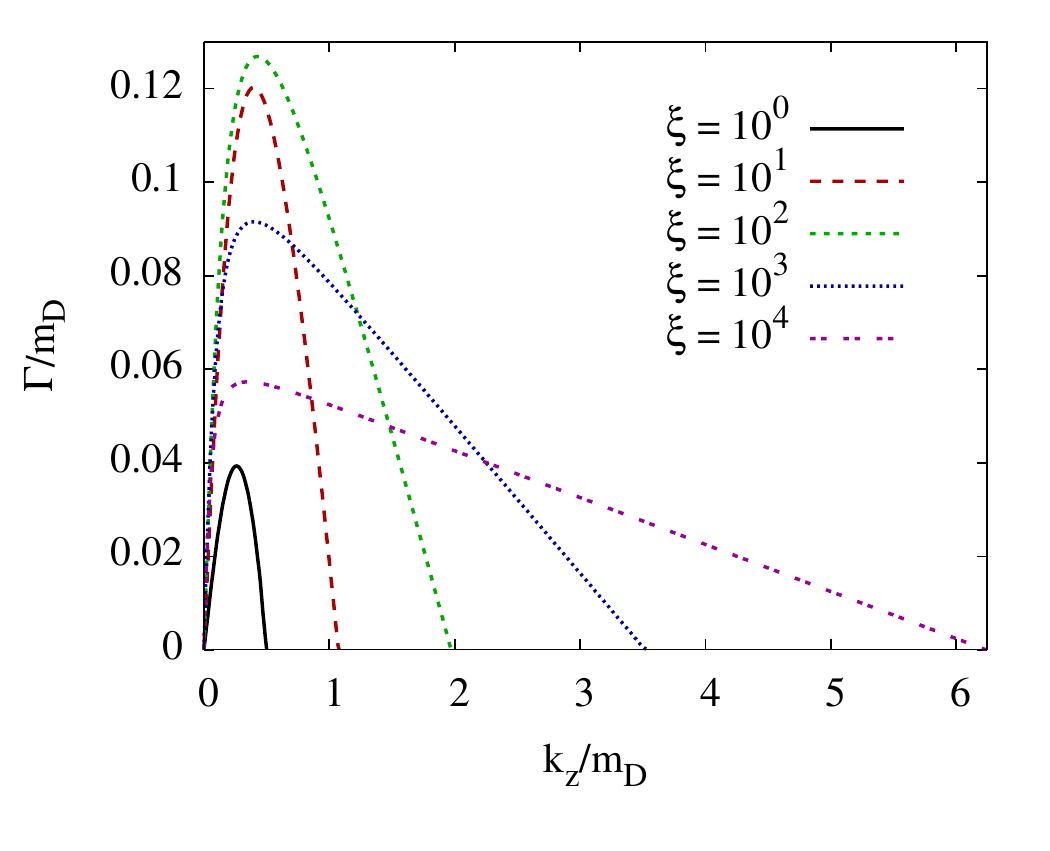}
\vspace{-7mm}
\caption{(Color online)
Unstable mode growth rate $\Gamma/m_D$ for fixed $\xi$ as a function of 
$k_z/m_D$ where $m_D$ is the Debye mass at the proper
time $\tau_{\rm iso}$. }
\label{fig:largexiGamma}
\end{figure}

In order to gain a qualitative understanding of the dynamics we can therefore focus our attention on the unstable mode 
spectrum for purely longitudinal modes.  In Fig.~\ref{fig:largexiGamma} we plot the unstable mode growth rate for
purely longitudinal modes for $\xi \in \{10^0,10^1,10^2,10^3,10^4\}$.  From this figure we can see that there
is a band of modes with positive unstable growth rate for longitudinal momenta $k_z \in (0,k_{z,\rm max})$ and 
there is a well-defined maximum growth rate $\Gamma^*$ at each value of $\xi$.  As $\xi$ increases $k_{z,\rm max}$ 
increases and for $\xi \gtrsim 10^2$ one finds that $\Gamma^*$ decreases monotonically.  This means that in an expanding 
plasma, more and more modes will become unstable as a function of proper time, but at the same time their growth rate 
is being reduced by the dilution of the plasma due to the 
longitudinal expansion.

It is possible to derive asymptotic relations for $k_{z,\rm max}$ and $\Gamma^*$ for large $\xi$ 
\cite{Rebhan:2005re}.  One finds that
\be
\lim_{\xi \gg 1} k_{z,\rm max} \sim m_D  (1+\xi)^{1/4} \, .
\ee
Using this we can determine the approximate proper time dependence of $k_{z,\rm max}$ for a longitudinal free 
streaming expansion
\be
\lim_{\tau \gg \tau_{\rm iso}} k_{z,\rm max} \sim m_D \left(\frac{\tau}{\tau_{\rm iso}}\right)^{1/2} \, .
\ee
Applying the same methodology to $\Gamma^*$ one finds
\be
\lim_{\tau \gg \tau_{\rm iso}} \Gamma^* \sim m_D(\tau) \sim m_D \left(\frac{\tau}{\tau_{\rm iso}}\right)^{-1/2} \, .
\ee
As a result, we can estimate the late time unstable growth by integrating $\Gamma^*$ to obtain
\bea
N(\tau) &\sim& 
\exp\left( m_D \int_{\tau_0}^{\tau} d \tau^\prime \left(\frac{\tau^\prime}{\tau_{\rm iso}}\right)^{-1/2} \right) \, ,
\nonumber \\ 
&\sim& \exp\left( 2 m_D \sqrt{\tau \tau_{\rm iso}} \right) \, ,
\eea
where we have suppressed an overall multiplicative constant.  We, therefore, see that the primary effect of longitudinal 
expansion will be to change the late time growth from being a pure exponential, as was the case in a static box, 
to $\exp(\sqrt{\tau})$.  To determine the precise nature of the dynamics on time scales relevant for heavy ion
collisions, however,  requires determining the full time evolution of all stable and unstable modes and properly 
taking into account their interactions.  We will now recall the derivation of the necessary dynamical equations to be solved numerically.


\section{Hard-Expanding-Loop Equations of Motion}
\label{sec:eom}

Our study is based upon the hard-loop approximation, which assumes
a separation of scales between the momenta of hard particles $p_s$ and
the momenta of soft collective fields $p_s \sim g\sqrt{f_h}p_h \ll p_h$ by
a sufficiently small gauge coupling $g$. This separation obviously requires
that $f_h$ is parametrically smaller than $1/g^2$. In an anisotropic
plasma, $f_h$ is moreover direction dependent and what actually
enters in the calculation of the parameters at the soft scale is
gradients $\partial f_{h}/\partial \mathbf p_h$. In terms of the
anisotropy parameter $\xi$ this means that at parametrically large $\xi$ the 
hard-loop approximation is applicable only as long as
 $\xi^{1/2}f_h$ is parametrically smaller than $1/g^2$. 

Because we are interested in investigating within the hard-loop framework
the earliest stages
of the evolution of a quark-gluon plasma, which according to the
CGC framework is born with overpopulated distribution functions
and with large anisotropy,
we shall treat the degree of anisotropy formally as being
of order 1 compared to $g$,
and $f_h$ of order $g^{-2+\epsilon}$. Eventually,
we boldly extrapolate our results to the very limits of
the hard-loop framework by setting $\epsilon=0$ and matching
with CGC parameters for the initial density and a strong
coupling $g$ that is numerically even larger than 
1.\footnote{In the notation of
Ref.~\cite{Kurkela:2011ub} where $f\sim \alpha_s^{-c}$,
$\tau\propto \alpha_s^{-a}$, $\xi^{1/2} \sim \delta^{-1}\equiv
\alpha_s^{-d}$, our framework is located at
parametric time $a=0$ with parametric occupancy $c=1-\frac{\epsilon}{2}$ and
parametric anisotropy $d=0$.} This matching to CGC parameters
is specified in Sec.\ \ref{sec:initialconditions-continuum}; in
the following we recapitulate the hard-expanding-loop equations,
which we have discussed in detail before in Ref.~\cite{Rebhan:2008uj},
and make the resulting equations explicit for the
case at hand, the fully (3+1)-dimensional evolution.

\subsection{Longitudinally expanding free streaming background solution}

In the hard-loop approximation,
the color neutral background distribution function $f_0(\mathbf p,\mathbf x,t)$
for the hard plasma particles has to satisfy
\be\label{vdf0}
v\cdot \partial\, f_0(\mathbf p,\mathbf x,t)=0,\qquad v^\mu=p^\mu/p^0.
\ee
This is trivially solved by a stationary distribution which only
depends on the momenta. Another solution is obtained by
considering a plasma with boost-invariant longitudinal expansion,
which we take as an approximation for the initial stage of
a heavy ion collision where the transverse extent of the system
is taken as sufficiently large.  
Assuming isotropy in transverse directions, $f_0$, which is
a Lorentz scalar, 
 can be written as
\cite{Baym:1984np,Mueller:1999pi}
\be
f_0(\mathbf p,x)=f_0(p_\perp,p^z,z,t)=f_0(p_\perp,p'^z,\tau)
\ee
where the Lorentz-boosted longitudinal momentum is
\be
p'^z=\gamma(p^z- \beta p^0),\;\;
\beta=z/t,\;\; \gamma=t/\tau,\;\; \tau=\sqrt{t^2-z^2},
\ee
with $p^0=\sqrt{p_\perp^2+(p^z)^2}$ for ultrarelativistic (massless) particles.

Switching to 
comoving coordinates
\bea
t = \tau \cosh\eta,\quad && \beta = \tanh\eta,\nn\\
z = \tau \sinh\eta,\quad  && \gamma = \cosh\eta,
\eea
we have curvilinear coordinates 
$ x^\alpha=(x^\tau,x^{i},x^\eta)=
(\tau,x^1,x^2,\eta)$ where here and elsewhere in the text indices $i,j,\ldots$ correspond 
to the two transverse spatial directions while Greek indices from the
beginning of the alphabet refer to the comoving spacetime coordinates. 
In these new coordinates
the metric reads 
\be\label{comovmetric}
ds^2=d\tau^2-d\mathbf x_\perp^2-\tau^2 d\eta^2=g_{\alpha\beta}(\tau)dx^\alpha
dx^\beta,
\ee
but
we shall continue to write our equations explicitly in terms of ordinary
derivatives and not deal with spacetime covariant derivatives. 
The gauge covariant derivative thus always 
means\footnote{The relation to 3-vectors is defined by $
\partial_\alpha=\partial/\partial  x^\alpha$ and
$A^\mu=(\phi,\vec A)$. 
Thus $ A_\alpha=(A_\tau,-A^x,-A^y,A_\eta)$.} 
$ D_\alpha=\partial_\alpha-ig[ A_\alpha,\cdot]$.

The field strength tensor is defined as 
$ F_{\alpha\beta}=\partial_\alpha  A_\beta
-\partial_\beta  A_\alpha
-ig[ A_\alpha, A_\beta]$ also in the comoving coordinates (with
all indices down), in which the
non-Abelian Maxwell equations can be written compactly as
\be
\frac{1}{\tau} D_\alpha(\tau  F^{\alpha\beta}) 
= j^\beta,
\ee
where indices have been raised with the inverse of the metric
$g_{\alpha\beta}(\tau)$
introduced in Eq.~(\ref{comovmetric}).

Similarly to space-time rapidity $\eta$, we define
momentum space rapidity $y$ for the massless particles according to
\be
p^\mu=p_\perp(\cosh y,\cos\phi,\sin\phi,\sinh y).
\ee
In comoving coordinates, this reads
\bea 
p^\tau&=&\sqrt{p_\perp^2+\tau^2( p^\eta)^2}\nn\\
&=&\cosh\eta\, p^0-\sinh\eta\, p^z
=p_\perp \cosh(y-\eta),\\
p^\eta&=&- p_\eta/\tau^2=(\cosh\eta\, p^z-\sinh\eta\, p^0)/\tau\nn\\
&=&p'^z/\tau
=p_\perp \sinh(y-\eta)/\tau.
\eea

Instead of the standard light-like vector $v^\mu=p^\mu/p^0$ 
which contains a unit 3-vector and which was
introduced in Eq.~(\ref{vdf0})
, we shall define 
\be
 V^\alpha = \frac{ p^\alpha}{p_\perp} =
\left(\cosh (y-\eta),\,\cos\phi,\,\sin\phi,\,\frac{1}{\tau}\sinh (y-\eta)\right),
\label{velocityDef1}
\ee
normalized such that it has a unit 2-vector in the transverse plane.

Since
\bea
 p^\tau \partial_\tau  p_\eta( x)\Big|_{y,\mathbf p_\perp}
&=&-p_\perp^2 \sinh(y-\eta)\cosh(y-\eta)\nn \, , \\ 
&=&- p^\eta\partial_\eta  p_\eta( x)\Big|_{y,\mathbf p_\perp} \, ,
\label{petaconstancy}
\eea
this can be solved by $f_0(\mathbf p,\mathbf x,t)=f_0(\mathbf p_\perp, p_\eta(x))
=f_0(\mathbf p_\perp,-p'^z(x)\tau(x))$.
For the case of longitudinal free streaming which is isotropic
at the particular 
proper time $\tau=\tau_{\rm iso}$ one can write $f_0$ in the form
\be\label{faniso}
f_0(\mathbf p,x)=f_{\rm iso}\!\left(\!\sqrt{p_\perp^2+(\frac{p'^z\tau}{\tau_{\rm iso}})^2}\!\right)\!
=f_{\rm iso}\!\left(\!\sqrt{p_\perp^2+ p_\eta^2/\tau_{\rm iso}^2}\!\right) .
\ee
Note that $f_0$ above falls into the general Romatschke-Strickland form for momentum-space 
anisotropic distribution functions \cite{Romatschke:2003ms}.

\subsection{Gauge-covariant Boltzmann-Vlasov  equations in a longitudinally expanding plasma}

In comoving coordinates the gauge-covariant Boltzmann-Vlasov equations
for colored perturbations $\delta f^a$ of a neutral collisionless plasma
with boost-invariant background distribution $f_0$
read
\be\label{VDf}
 V\cdot  D\, \delta f^a\big|_{p^\mu}=g  V^\alpha 
 F_{\alpha\beta}^a  \partial_{(p)}^\beta f_0(\mathbf
p_\perp, p_\eta).
\ee
Here the derivative on the left-hand side has to be taken at fixed 
Cartesian $p^\mu$
rather than fixed comoving $p^\alpha$. Notice also that only derivatives
of $f_0(\mathbf
p_\perp, p_\eta)$ with $\partial_{(p)}^\beta$ where the 4-index is up
do not introduce explicit $\tau$ dependence so that one still
has $p\cdot\partial \,( \partial_{(p)}^\beta f_0)|_p=
 p\cdot\partial \,( \partial_{(p)}^\beta f_0)|_p=0$.

Eq.~(\ref{VDf}) can be solved in terms of an auxiliary field
$ W_\beta( x;\phi,y)$ 
that does not depend
on the hard scale $p^0$ and which is defined by
\be\label{Wdef}
\delta f(x;p)=-g  W_\beta( x;\phi,y) 
\partial_{(p)}^\beta f_0(p_\perp, p_\eta),
\ee
if it
satisfies
\be\label{VDW}
 V\cdot  D\,  W_\beta\big|_{\phi,y}= V^\alpha  F_{\beta\alpha} \, .
\ee

Since the fluctuations $\delta f^a$ give the induced current in
\be\label{jind}
D_\mu F^{\mu\nu}_a=j^\nu_a=
g\, t_R \int{d^3p\over (2\pi)^3} \frac{p^\mu}{2 p^0} \delta f_a(\mathbf p,\mathbf x,t),
\ee
$j$ can be expressed in terms of integrals over the $W$ fields.
(Here $t_R$ is a suitably normalized group factor, while the total
number of degrees of freedom of the hard particles is contained in 
the normalization of the distribution function $f_0$.)

With (\ref{faniso})
we have
\bea
\partial_{(p)}^\beta f_0&=&f_0'\partial_{(p)}^\beta
\sqrt{p_\perp^2+ p_\eta^2/\tau_{\rm iso}^2}\nn\\
&=& f_0' \, 
\frac{\left(0,-\cos\phi,-\sin\phi,-\frac{\tau}{\tau_{\rm iso}^2}\sinh(y-\eta)\right)
}{\sqrt{1+\frac{\tau^2}{\tau_{\rm iso}^2}\sinh^2(y-\eta)}} \, , \nonumber \\
\eea
which yields
\bea
 j^\alpha &=& -\frac{m_D^2}{2}\!\int_0^{2\pi} \frac{d\phi}{2\pi}
\int_{-\infty}^\infty dy \,  V^\alpha
\nonumber \\
&& \times \!\left(\! 1+\frac{\tau^2}{\tau_{\rm iso}^2}\sinh^2(y-\eta) \!\right)^{\!-2} \!\! \mathcal W( x;\phi,y) \, ,
\label{current}
\eea
where
\bea
&&\mathcal W =  V^{i} W_{i}
-\frac{1}{\tau_{\rm iso}^2} V_\eta\,  W_\eta,\qquad \nn\\ &&
 V^{i}=(\cos\phi,\sin\phi) \, ,\quad
 V_\eta = -\tau \sinh(y-\eta) \, ,
\eea
and
\be
m^2_D=-g^2t_R \int_0^\infty \frac{dp\,p^2}{(2\pi)^2} f'_{\rm iso}(p) \, .
\ee
The (constant) mass parameter $m_D$ equals the Debye mass at the proper time $\tau_{\rm iso}$.

The combination $\mathcal W$ introduced above satisfies
\bea
\label{VDtildeW}
 V\cdot  D\, \mathcal W
&=& \left( V^{i}  F_{i\tau}+\frac{\tau^2}{\tau_{\rm iso}^2} V^\eta  F_{\eta\tau}\right) V^\tau 
\nonumber\\
&&+ V^{i} V^\eta  F_{i\eta}\left(1-\frac{\tau^2}{\tau_{\rm iso}^2}\right).
\eea
This single equation for $\mathcal W$ in combination 
with the Yang-Mills equations
and the integral giving $j$ in terms of $\mathcal W$ 
closes our equations of motion. To solve them
numerically, we adopt the comoving
temporal gauge $A^\tau=0$ and introduce canonical conjugate field momenta
for the remaining gauge fields according to
\be
\Pi^{i} = \tau \partial_\tau A_{i} = 
-\tau \partial_\tau A^{i}=-\Pi_{i} \, ,
\ee
and
\be
\Pi^\eta=\frac{1}{\tau}\partial_\tau A_\eta \, .
\ee

In terms of fields and conjugate momenta,
the Yang-Mills equations take the form
\bea\label{tagYM}
\tau \partial_\tau \Pi^\eta&=&j_\eta-D_{i} F^{i}{}_\eta\,,\\
\tau^{-1}\partial_\tau \Pi_{i}&=&j^{i}
-D_{j}F^{{j}{i}}
-D_\eta F^{\eta{i}}\,,
\eea
while the Gauss law constraint takes the form
\be
\tau j^\tau = D_\eta \Pi^\eta + D^i\Pi_i \, .
\label{gausslawcontinuum3d}
\ee

In temporal
gauge, where 
$F_{i \tau} =  \Pi_i/\tau$ and $F_{\eta \tau} = - \tau \Pi^{\eta}$,
the field equation for $\mathcal W$, Eq.\ (\ref{VDtildeW}),
 becomes
\bea
\label{DVW3+1}
\partial_\tau \mathcal W(\tau,{\bf x}_\perp,\eta;\phi,y) &=& 
-\frac{1}{\cosh(\bar y)} \left[ v^{i} D_{i} {\mathcal W} + \frac{\sinh(\bar y)}{\tau} D_{\eta} {\mathcal W} \right]
\nonumber \\
&& +\frac{1}{\tau}v^{i} \Pi_{i} -\frac{\tau^2 \sinh(\bar y) }{ \tau_{\rm iso}^2} \Pi^\eta
\nonumber \\
&& + \frac{\tanh(\bar y)}{\tau} \left(1- \frac{\tau^2}{\tau_{\rm iso}^2} \right)
v^{i} F_{{i}\eta}
\,,\\
\mbox{with}&& \bar y\equiv y-\eta.\nonumber
\eea
In the limit that all fields are independent of the transverse spatial directions Eqs.\ (\ref{tagYM})--(\ref{DVW3+1})
reduce to the 1D+3V equations of Ref.~\cite{Rebhan:2008uj}.

We can recast (\ref{DVW3+1}) into a form which is more convenient for computing the currents
in Eq.~(\ref{current}) by defining
\be
\mathcal W(\tau,{\bf x}_\perp,\eta;\phi,y) \equiv
 \bar f(\tau,\tau_{\rm iso},\bar y) \, \overline{\mathcal W}(\tau,{\bf x}_\perp,\eta;\phi,\bar y)  \, ,
\ee
with
\be
\bar f(\tau,\tau_{\rm iso},\bar y) = \left(1+\frac{\tau^2}{\tau_{\rm iso}^2}\sinh^2\bar y\right)^2 \, .
\ee
We also replaced $y$ by $\bar y\equiv y-\eta$ as argument of $\overline{\mathcal W}$ because the auxiliary fields turn out to be peaked around $y\sim \eta$.

Now using
\bea
\partial_\tau {\mathcal W} &=& \bar f \, \partial_\tau \overline{\mathcal W} + \frac{\partial \bar f}{\partial \tau} \, \overline{\mathcal W} \\
D_\eta {\mathcal W}(\tau,{\bf x}_\perp,\eta;\phi,y) &=& 
(D_\eta - \partial_{\bar y}) \left[ \bar f \overline{\mathcal W}(\tau,{\bf x}_\perp,\eta;\phi,\bar y) \right]
\nonumber \\
	&=& \bar f\, (D_\eta - \partial_{\bar y})\overline{\mathcal W} - 
\frac{\partial \bar f}{\partial \bar y} \, \overline{\mathcal W} \\
D_i {\mathcal W} &=& \bar f D_i \overline{\mathcal W} \, , 
\eea
together with
$$
{\rm tanh}\,{\bar y} \; \frac{\partial \bar f}{\partial \bar y} = \frac{\partial \bar f}{\partial \tau} \, ,
$$ 
we obtain
\bea
\label{DVW3+1v2}
\partial_\tau \overline{\mathcal W}(\tau,{\bf x}_\perp,\eta;\phi,\bar y) &=& 
-\frac{1}{{\rm cosh}\,\bar{y}} \biggl[ v^{i} D_{i} \overline{\mathcal W} \nonumber\\
&&+ \frac{{\rm sinh}\,\bar{y}}{\tau} \left( D_\eta\overline{\mathcal W} 
- \partial_{\bar y}\overline{\mathcal W} \right) \biggr]
\nonumber \\ 
+ \frac{1}{\bar f(\tau,\tau_{\rm iso},\bar y)} \!\! && \!\!
\biggl[ \frac{1}{ \tau}v^{i} \Pi_{i} -\frac{\tau^2 {\rm sinh}\,\bar{y} }{ \tau_{\rm iso}^2} \Pi^\eta 
\nonumber \\ &&  
+ \frac{{\rm tanh}\,\bar{y}}{\tau} \left(1- \frac{\tau^2}{\tau_{\rm iso}^2} \right)
v^{i} F_{{i}\eta} \biggr]
\, .\qquad\quad
\eea
In terms of $\overline{\mathcal W}(\tau,{\bf x}_\perp,\eta;\phi,\bar y)$ the expression for 
the current (\ref{current}) simplifies to
\bea
&& j^\alpha(\tau,{\bf x}_\perp,\eta) = \nonumber\\
&&\quad - \, \frac{m_D^2}{2}\int_0^{2\pi} \frac{d\phi }{ 2\pi}
\int_{-\infty}^\infty d{\bar y} \;  V^\alpha \,
\overline{\mathcal W}(\tau,{\bf x}_\perp,\eta;\phi,\bar y) \, .\qquad
\label{currentv2}
\eea
%
%
The equations of motion listed above are numerically solved by discretizing them in space and 
velocity space (hence the designation
3D+3V). The gauge fields live on the 3-dimensional space
parametrized by space-time rapidity $\eta$ and two transverse coordinates $\mathbf x_\perp$. The $\mathcal W$ field lives additionally in velocity space, which
because of the masslessness of the hard particles is, in the end, 2-dimensional, parametrized by $\bar y$ and $\phi$.

For the details of the lattice discretizations used we refer the reader to 
Appendix \ref{app:eom-lattice}.


\section{Initial Conditions}
\label{sec:initialconditions-continuum}

\subsection{Matching of the Debye mass with CGC parameters}

As in our 1D+3V simulations \cite{Rebhan:2008uj},
we evolve from an initial time $\tau_0 \simeq Q_s^{-1}$
and fix the density of our initial plasma such that it matches
estimates obtained from the CGC framework.

According to Ref.~\cite{Baier:2002bt}, the initial hard-gluon density
can be written as
\be\label{hardgluonstau0}
n(\tau_0)=c \frac{N_g Q_s^3}{4\pi^2 N_c\alpha_s (Q_s \tau_0)},
\ee
with $c$ being the gluon liberation factor, which following
an analytical estimate by Kovchegov \cite{Kovchegov:2000hz} we choose
as $c=2\ln2\approx 1.386$. While being significantly higher
than the original estimates $c\simeq 0.5$ of
Ref.~\cite{Krasnitz:2001ph,Krasnitz:2003jw}, this value
is in fact rather close to the most recent numerical result $c\simeq 1.1$
by Lappi \cite{Lappi:2007ku}.

In our effective field equations, the initial hard-gluon density
enters only through the mass parameter $m_D$, which is defined
as the Debye mass at the proper time $\tau_{\rm iso}$.
In the glasma phase of the CGC framework, the pressure at
early times is strongly anisotropic, with the longitudinal
pressure starting out even with negative values.
To model this approximately, we formally choose $\tau_{\rm iso}\ll \tau_0$,
so that our initial particle distribution has initial pressure 
$P_L \ll P_T$. Sticking to our previous choice in Ref.~\cite{Rebhan:2008uj}
we take $\tau_{\rm iso}=0.1 \, \tau_0$.
The correspondingly oblate distribution function is
taken to be obtained from
$f_{\rm iso}(p)=\mathcal N (2N_g) / (e^{p/T}-1)$, where $N_g=N_c^2-1$ 
is the number
of gluons,
since in CGC calculations an approximately thermal distribution
was obtained for the gluon distribution in transverse directions. Following 
Ref.~\cite{Iancu:2003xm} we set this transverse temperature $T=Q_s/d$ with $d^{-1}\simeq 0.47$.
Eq.~(\ref{hardgluonstau0}) then fixes the normalization factor $\mathcal N$
through
\be
n(\tau_0)\frac{\tau_0}{\tau_{\rm iso}}=
n(\tau_{\rm iso})=\frac{2 \zeta(3)}{\pi^2}\mathcal N N_g T^3.
\ee
In a plasma containing only gluons with distribution function
$f_{\rm iso}$, the Debye mass is given by
\be
m_D^2(\tau_{\rm iso})=\mathcal N \frac{4\pi \alpha_s N_c T^2}{3}.
\ee
With $N_c=3$ and the above values for $c$ and $d$ we thus obtain
\be\label{mD2CGC}
m_D^2(\tau_{\rm iso})\tau_0^2(Q_s \tau_0)^{-1}= \frac{\pi c d }{6 \zeta(3)}
\frac{\tau_0}{\tau_{\rm iso}} 
\approx 1.285
\frac{\tau_0}{\tau_{\rm iso}} .
\ee
In our previous studies of a stationary anisotropic
plasma we have observed little difference between
simulations using gauge group SU(2) versus
SU(3) provided the same value of $m_D$ was used
\cite{Rebhan:2005re,Ipp:2010uy}, so we adopt the
value (\ref{mD2CGC}) for our simulations with gauge group SU(2).

Notice that in the above matching which involved an overpopulated
distribution function $n(\tau_0)\propto \alpha_s^{-1}$
the gauge coupling dropped out in the mass parameter $m_D^2$.
As discussed in Section \ref{sec:eom}, this means that we
are extrapolating the hard loop framework, which assumes
a parametric separation of hard and soft scales, to its very limits.
In the following we shall compare hard and soft contributions
to the pressure and find that the soft field contributions
are small compared to the hard particle contributions
even after plasma instabilities have grown nonperturbatively strong.
As long as this is the case, we assume that the hard loop framework
is still applicable.

In order to compare soft and hard contributions, we finally have
to fix the gauge coupling. For that purpose we shall choose
$\alpha_s=0.3$ or $g=1.94$ as a representative value.

\subsection{Initial field fluctuations}

In order to have seed fields for the unstable modes in an
anisotropic plasma with oblate anisotropy, initial fluctuations
that break perfect boost invariance are required.
Fluctuations in the sources of heavy-ion collisions as well
as vacuum fluctuations in all fields are inevitable, and
by ``natural selection'' those fluctuations which lead to
the most rapid onset of growth will dominate all later dynamics.

In previous hard-loop lattice simulations with fixed anisotropy
the question of which initial conditions to choose was rather unimportant
as long as unstable modes were excited. 
Seed fields in chromo-fields or in $\mathcal W$
fields were considered on the basis of convenience.

As it turns out,
more care is needed in the expanding case. In \cite{Romatschke:2006wg},
where the formalism of hard expanding loops was introduced and
studied semi-analytically in the (1+1)-dimensional Abelian case, only 
initial conditions formulated
in terms of transverse electric fields were considered. Likewise,
only seed fields in transverse chromo-fields
were subsequently employed in the numerical 1D+3V non-Abelian lattice
study of Ref.~\cite{Rebhan:2008uj}, which in the weak-field regime
reproduced the earlier semi-analytical results, and thus also
the original finding of a (with regard to heavy-ion collisions)
uncomfortably long delay of the onset
of growth of plasma instabilities. (The generalization considered
in Ref.~\cite{Rebhan:2008uj}, namely to also intialize magnetic fields
did not change this conclusion.)

In Ref.~\cite{Rebhan:2009ku} the semi-analytical treatment
of Ref.~\cite{Romatschke:2006wg} was generalized to the much
more complex case of generic
(3+1)-dimensional Abelian modes in an expanding plasma,
and at this occasion also the most general initial conditions were
considered, involving both electric and magnetic fields as well as the
auxiliary $\mathcal W$
fields which describe fluctuations in the induced currents.
Surprisingly enough, initial fluctuations in the $\mathcal W$ fields
lead to a drastic (order-of-magnitude) reduction of the initial delay of the onset of growth.
Evidently, initial conditions in the electric
and magnetic fields predominantly give stable plasmon modes
and less strongly excite the unstable modes. The latter are
instead more easily triggered by fluctuations in the induced currents
described by the $\mathcal W$ fields.

The simplest initial conditions that provide seed fields for
Weibel instabilities while having initial vanishing charge density
are $\phi$ and $y$-independent fluctuations of the component
fields $W_i(x;\phi,y)$ and $W_\eta(x;\phi,y)$.
The former induce transverse currents which are most directly
related to the $\alpha$ modes, whereas a $\phi$ and $y$-independent 
$W_\eta$ seeds longitudinal
currents that are less important for the plasma instabilities.
Because of their subdominant effect, we have mostly omitted
$W_\eta$ seeds and only kept $W_i(x;\phi,y)$ when
assembling the initial $\mathcal W$ field.

Another point to consider is the spectrum of initial fluctuations.
Because we are using highly anisotropic lattices with particularly
fine resolution in the longitudinal direction, initializing with
white noise fluctuations would correspond to very high UV noise
in longitudinal wave numbers. We have therefore implemented an
adjustable mode number cutoff, $\Lambda_\nu$, in wave numbers $\nu$ 
dual to the rapidity variable $\eta$ and 
populate all modes $\propto e^{i\nu\eta}$
equally below this cutoff, with white noise in transverse
directions. Because the ``natural selection'' of plasma instabilities
quickly picks out the most strongly growing modes, we have refrained
from attempts to model the initial spectrum other than
ensuring that a good range of seeds is available.


\section{Observables}
\label{sec:observables}

Here we list the quantities which we will present in the results section.  We present only the
continuum formulae.  For the details of the lattice discretizations used we refer the reader to 
Appendix \ref{app:eom-lattice}.  Note that in most of the results presented we have averaged 
observables over a set of runs in order to account for variations in the random initial conditions 
employed. 

\subsection{Field energy densities and pressures}

The transverse/longitudinal electric and magnetic components of the field energy density are given by
\bea
\mathcal E=\mathcal E_T+\mathcal E_L &=&
\mathcal E_{B_T}+\mathcal E_{E_T}+\mathcal E_{B_L}+\mathcal E_{E_L}\nn\\
&=&\tr\left[ \tau^{-2} F_{\eta i}^2 +
\tau^{-2}\Pi_{i}^2 +
F_{xy}^2+ \left(\Pi^{\eta}\right)^2 \right] ,
 \nonumber \\
\label{eq:edens}
\eea
and the Hamiltonian density is given by $\mathcal H=\tau\mathcal E$.  
The transverse and longitudinal field pressures are obtained via
\bea
\mathcal P_L^{\rm field} &=& {\mathcal E}_T - {\mathcal E}_L \, , \\
\mathcal P_T^{\rm field} &=& {\mathcal E}_L \, .
\label{eq:fieldpressures}
\eea
Note that from the above one has at all times $2 \mathcal P_T^{\rm field} + \mathcal P_L^{\rm field} = \mathcal E$ such
that the energy momentum tensor is traceless.

\subsection{Particle Pressures}

In a comoving frame,
the energy density and pressure components of the hard particle background
can be determined by evaluating 
\begin{align}
T^{\alpha\beta}_{\rm part.}=
(2\pi)^{-3}\int d^2p_T dy\, p^\alpha p^\beta f_0 \,,
\end{align}
which yields
\bea
&&\mathcal E_{\rm part.}(\tau)=T^{\tau\tau}_{\rm part.} \nonumber \\
&& \hspace{8mm}
= \frac1{2}\left[\frac1{{\bar\tau}^2}+\frac{\arcsin\sqrt{1-{\bar\tau}^{-2}}}{\sqrt{{\bar\tau}^2-1}}\right]\mathcal E_{\rm iso} ,
\\
&&\mathcal P_T^{\rm part.}(\tau)=\frac12 T^{ii}_{\rm part.} \nonumber \\
&&\hspace{8mm}
= \frac1{4({\bar\tau}^2-1)}
\left[1+\frac{{\bar\tau}^2-2}{\sqrt{{\bar\tau}^2-1}}\arcsin\sqrt{1-{\bar\tau}^{-2}} \right]\mathcal E_{\rm iso},
\nonumber \\ &&
\label{eq:particlePT}
\\
&&\mathcal P_L^{\rm part.}(\tau)=-T^\eta_{{\rm part.}\eta}\nonumber \\
&&\hspace{8mm}
= \frac1{2({\bar\tau}^2-1)} 
\left[-\frac1{{\bar\tau}^2} +  \frac{\arcsin\sqrt{1-{\bar\tau}^{-2}}}{\sqrt{{\bar\tau}^2-1}} \right]\mathcal E_{\rm iso},
\label{eq:particlePL}
\eea
where $\mathcal E_{\rm iso}=\mathcal E_{\rm part.}(\tau_{\rm iso})$,
$\bar \tau \equiv \tau/\tau_{\rm iso}$ and we have assumed $\bar \tau \ge1$. \\

In the results section as a measure of isotropization we will present plots of the ratio 
\be
\frac{\mathcal P_L}{\mathcal P_T} = \frac{\mathcal P_L^{\rm field} + \mathcal P_L^{\rm part.}}
{\mathcal P_T^{\rm field} + \mathcal P_T^{\rm part.}} \, .
\label{eq:totalpressure}
\ee
If this quantity is less that one, then the system possesses an overall oblate momentum-space
anisotropy and if it is greater than one, then it possesses a prolate momentum-space anisotropy.

\subsection{Energy spectra}

In order extract spectral information about the field configurations, the canonical way to 
proceed is to gauge fix to a spatially smooth gauge such as Coulomb gauge and then extract
mode occupation numbers from either the electric or magnetic fields \cite{Mandula:1987rh,%
Moore:1997cr,Arnold:2005ef}.  However, such a method is not free from ambiguity in the 
infrared due to the lingering problem of large gauge transformations (aka Gribov copies).  

Here we follow a different method introduced by Fukushima and Gelis \cite{Fukushima:2011nq}
in which we extract the electric and magnetic fields at a given proper time from the lattice
simulation using
\bea
E_i({\bf x}_T,\eta) &=& \tau^{-1} \Pi_i  \, , \nonumber \\
E_L({\bf x}_T,\eta) &=& \Pi^\eta \, , \nonumber \\
B_x({\bf x}_T,\eta) &=& F_{\eta y} \simeq \frac{2}{i g a_\eta \tau} \tr[t^a(1-U_{\eta y})] \, , \nonumber \\
B_y({\bf x}_T,\eta) &=& F_{\eta x} \simeq \frac{2}{i g a_\eta \tau} \tr[t^a(1-U_{\eta x})] \, , \nonumber \\
B_L({\bf x}_T,\eta) &=& F_{xy} \simeq \frac{2}{i g} \tr[t^a(1-U_{xy})] \, . 
\eea
We then perform a three-dimensional Fourier transform of each field component, e.g.
\be
E_i({\bf k_T},\nu) = \int \frac{d^2x_T}{(2\pi)^2} \frac{d\eta}{2\pi}
E_i({\bf x_T},\eta) e^{i {\bf k}_T \cdot {\bf x}_T} e^{i \nu \eta} \, .
\label{eq:eFT}
\ee
Since we are primarily interested in the longitudinal spectra, we integrate over the transverse 
wavevectors to obtain, e.g.
\be
E_i(\nu) = \int \frac{d^2 k_T}{(2 \pi)^2} \, E_i({\bf k_T},\nu) \, . \nonumber \\
\label{eq:eFTnu}
\ee

Having obtained the field components we can decompose the energy
density in terms of the longitudinal wavenumber
\bea
{\mathcal E}_E &=& \int \frac{d\nu}{2 \pi} {\mathcal E}_E(\nu)  =  \int \frac{d\nu}{2 \pi} [ {\mathcal E}_{E_L}(\nu) + {\mathcal E}_{E_T}(\nu) ]\, ,
\nonumber \\
{\mathcal E}_B &=& \int \frac{d\nu}{2 \pi} {\mathcal E}_B(\nu)  =  \int \frac{d\nu}{2 \pi} [ {\mathcal E}_{B_L}(\nu) + {\mathcal E}_{B_T}(\nu) ]\, ,
 \nonumber \\
\label{eq:nusum}
\eea
where we have the energy density at each longitudinal wavenumber
\bea
{\mathcal E}_{E_L}(\nu) &=& \tr[E_L(-\nu) E_L(\nu)] \nonumber = \tr|E_L|^2 \, , \nonumber \\
{\mathcal E}_{E_T}(\nu) &=&  \sum_{i\in\{x,y\}}\tr[E^i(-\nu) E^i(\nu)] = \sum_{i\in\{x,y\}} \tr|E^i|^2 \, , \nonumber \\
{\mathcal E}_{B_L}(\nu) &=& \tr[B_L(-\nu) B_L(\nu)] \nonumber = \tr|B_L|^2 \, , \nonumber \\
{\mathcal E}_{B_T}(\nu) &=&  \sum_{i\in\{x,y\}}\tr[B^i(-\nu) B^i(\nu)] = \sum_{i\in\{x,y\}} \tr|B^i|^2 \, ,
 \nonumber \\
\label{eq:ennu}
\eea
where the traces are color traces.  
The total longitudinal energy spectra are obtained by summing all components
\be
{\mathcal E}(\nu) = {\mathcal E}_{E_L}(\nu) + {\mathcal E}_{E_T}(\nu)  + {\mathcal E}_{B_L}(\nu)  + {\mathcal E}_{B_T}(\nu)  \, . 
\label{eq:longenergyspec}
\ee

The spectral decomposition (\ref{eq:ennu}) is not gauge invariant; gauge transformations could in principle still redistribute 
the energy distribution in $\nu$, but this redistribution is limited by the fact that the integrals (\ref{eq:nusum}) is gauge 
invariant. We thus expect that the degree of gauge dependence is much milder than in bare mode occupation numbers of 
the gauge fields before they are made maximally smooth by going to Coulomb gauge.

Note that one can compute the total energy density via Eq.~(\ref{eq:edens}) and Eq.~(\ref{eq:nusum}) and compare as 
a crosscheck of the spectra calculation.  Numerically we find very good agreement between the two methods.  We have also 
performed a Fourier analysis of the spatial distribution of the (gauge-invariant) chromo-field energy on the lattice. Besides 
the expected peak at zero momentum, we found that the remaining spatial fluctuations reflect closely the spectral 
decomposition defined through Eqs.~(\ref{eq:ennu}).

\section{Results}
\label{sec:results}

In this section we present the results of our numerical simulations for SU(2) gauge fields which include:  real-time 
gauge field energy densities, particle and field pressures, energy spectra, and fit to the energy spectra.  For all results 
shown in this section we initialize current fluctuations  (via $\mathcal W$ fields) with an amplitude $\Delta$ as described in Section 
\ref{sec:initialconditions-continuum} and Appendix \ref{app:eom-lattice-initialcondition}.  In order to generate 
occupation numbers $\sim 1/2$ consistent with those expected from initial quantum-mechanical rapidity fluctuations 
\cite{Fukushima:2006ax} one should choose $\Delta \sim 1.6$.  Unfortunately, due to numerical limitations stemming
from the fact that we simulate compact gauge groups, we are unable to use such a large value of $\Delta$.  Instead in the 
main plots shown below we use an initial current fluctuation amplitude of $\Delta = 0.8$ which can be expected to result in longer 
isotropization times than one would obtain with the larger seed values necessary.  In order to assess the dependence 
of our results on $\Delta$ we present the variation of the energy density and pressure ratio.  In the conclusions we will
discuss the extrapolation of our result to realistic values of $\Delta$.

\begin{figure}[t]
\includegraphics[width=0.5\textwidth]{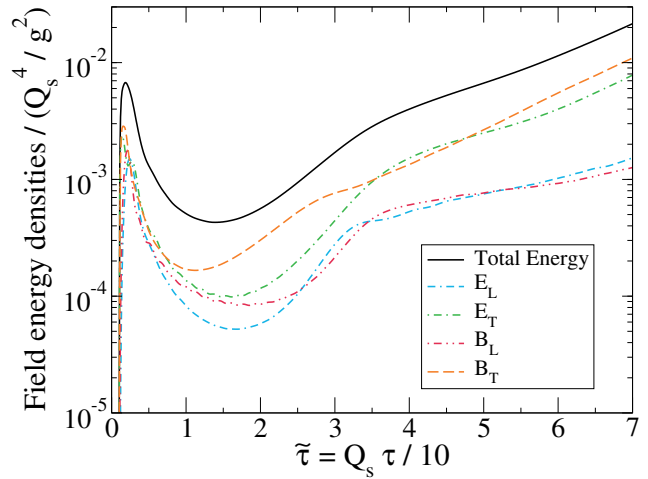}
\caption{(Color online)
Chromoelectric, chromomagnetic, and total energy densities (\ref{eq:edens}) as a function of proper time
from averaging over our standard set of runs.  Proper time is normalized such that when using $Q_s = 2$ GeV
each unit of $\Delta\tilde\tau$ is 1 fm/c.  See text for simulation parameters used.}
\label{fig:eDensity1}
\end{figure}

For all results shown the lattice spatial size was $N_T^2 \times N_\eta = 40^2 \times 128$ with transverse 
lattice spacing of $a = Q_s^{-1}$ and longitudinal lattice spacing of $a_\eta = $~0.025.  The lattice size in velocity 
space was $N_u \times N_\phi = 128 \times 32$.  The longitudinal spectral cutoff for the current-based rapidity 
fluctuations was taken to be $\Lambda_\nu \nu_{\rm min} = 8 \nu_{\rm min}
\approx 15.7$.  The initial time was taken to be $\tau_0 = Q_s^{-1}$ and we 
used $\tau_{\rm iso}/\tau_0=0.1$.  For the temporal time step we use $\epsilon = 10^{-2}\tau_0$.  For 
details of the lattice discretizations used for the equations of motion we refer the reader to Appendix 
\ref{app:eom-lattice}.  When plotting observables as a function of time we will plot them as a function of $\tilde\tau
\equiv Q_s \tau/10$.  For LHC one has $Q_s \simeq 2$~GeV = (0.1 fm)$^{-1}$ and for RHIC one has $Q_s \simeq 1.4$~GeV = 
(0.14 fm)$^{-1}$.  The division by a factor of 10 makes it so that when considering LHC energies each interval of
$\Delta \tilde\tau = 1$ is 1 fm/c.  At RHIC each interval of $\Delta \tilde\tau = 1$ is 1.4 fm/c.

For numerical tests such as varying the lattice spacing, lattice size, spectral cutoffs, and velocity resolution we refer the reader 
to App.~\ref{app:numerics}.
The lattice equations of motion are written in terms of rescaled dimensionless fields.  When comparing pressures in 
soft fields with pressures from hard particles, we have assumed a value of $g = 1.94$ consistent with $\alpha_s=0.3$
which is in the right ball park for RHIC and LHC heavy ion collisions.  
Note that formally our results are only trustable in the weak-coupling limit and we are making a bold extrapolation
when we assume $\alpha_s=0.3$.  Nevertheless, we do this in order to obtain a rough estimate of the
isotropization time associated with the chromo-Weibel instability in a background which is undergoing longitudinal 
free-streaming expansion.

\subsection{Energy densities}

\begin{figure}[t]
\includegraphics[width=0.5\textwidth]{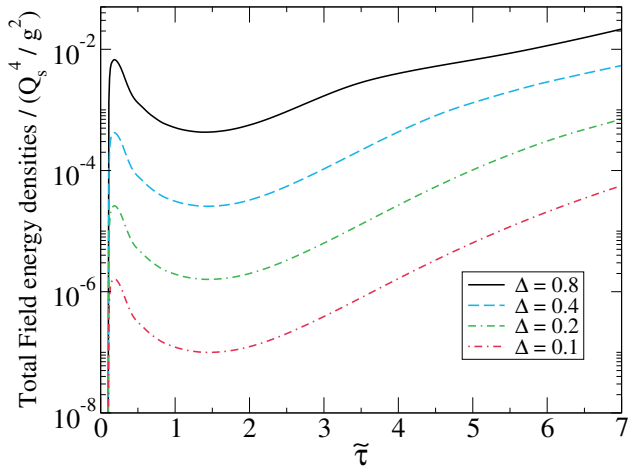}
\caption{(Color online) 
Total field energy density for different initial current fluctuation magnitudes $\Delta \in \{0.1,0.2,0.4,0.8\}$.  
See text for simulation parameters used.
}
\label{fig:eDensity2}
\end{figure}

In Fig.~\ref{fig:eDensity1} we show the chromoelectric, chromomagnetic, and total energy densities (\ref{eq:edens}) 
as a function of proper time.  The results shown are averaged over 50 runs which will serve as our standard set of runs
for most observables in this section.\footnote{In App.~\ref{app:numerics} Fig.~\ref{fig:numtests-collection-a} we plot the total
field energy density resulting from all 50 runs for comparison.}
From Fig.~\ref{fig:eDensity1} we see that for the first $\tilde\tau \lesssim$
1.2 the soft fields are depleted by the longitudinal expansion.  After this time the unstable modes present in the
initial condition begin to show appreciable growth.  Initially all components of the chromofield start out with 
approximately equal energy density, but at this time the system begins to be dominated by transverse chromomagnetic
fields.  However, due to the large amplitude of the initial current fluctuations we quickly see the development of large
transverse chromelectric fields followed by rapid growth in the longitudinal chromoelectric and chromomagnetic fields.

All field components become approximately the same magnitude at a time of $\tilde\tau \sim $~3.5
when $\Delta=0.8$.  We will refer to the point
in time at which all components of the field energy density give approximately the same contribution as the 
``non-Abelian point''.  From this point on, in
contrast to the fixed-anisotropy simulations, one does not see a saturation of the exponential growth, just a moderate 
reduction of the growth rate.  Instead we see that,
similar to the 1D+3V simulations, the transverse chromoelectric and chromomagnetic fields begin to dominate the
energy density and do so for the rest of the simulation.  As we will see below, by the end of the simulation a large
portion of the energy is in ultraviolet longitudinal lattice modes and one starts to see lattice artifacts; however, up to 
this point we see no sign of saturation of the roughly exponential growth in the chromofields.

In Fig.~\ref{fig:eDensity2} we show the total field energy density for different initial current fluctuation amplitudes 
$\Delta \in \{0.1,0.2,0.4,0.8\}$.  As can be seen from this figure, apart from a slight reduction in unstable mode growth
when the fields reach the non-Abelian point (which
moves to large times for smaller $\Delta$), the behavior is qualitatively independent of the assumed amplitude.  We 
note that there is a fundamental limit on how large one can make $\Delta$ without violating the assumptions of the 
hard-loop effective theory we employ.  In practice, this limit is set by the physical requirement that the majority
of the energy density should still be contained in the hard particle distribution function.  We note that for $\Delta=0.8$
we are still safely below this bound with the initially induced soft fields only carrying only $\sim$ 1.5\% of the total
energy with the vast majority of the energy coming from the hard sector.

\subsection{Pressures}

\begin{figure}[t]
\includegraphics[width=0.5\textwidth]{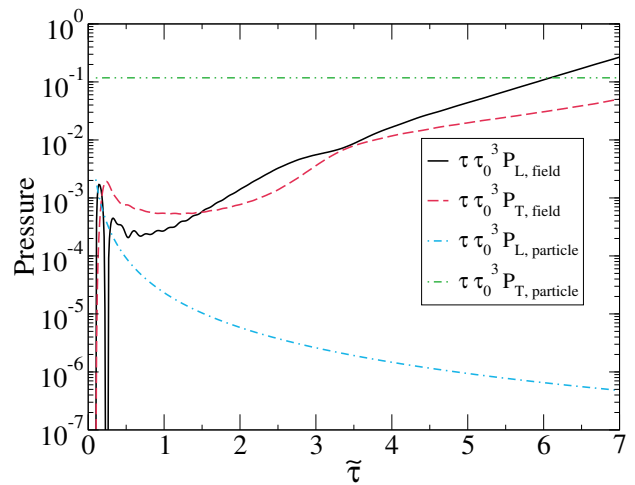}
\caption{(Color online) 
Hard particle and field pressures scaled by $\tau_0^3 \tau$ as a function of proper time.  The data were taken from the 
same set of runs as Fig.~\ref{fig:eDensity1}.
}
\label{fig:pressure1}
\end{figure}
 
In Fig.~\ref{fig:pressure1} we show the hard particle and field pressures scaled by $\tau_0^3 \tau$ as a function of proper 
time.  The data were taken from the same set of runs as Fig.~\ref{fig:eDensity1} and the pressures were computed
using Eq.~(\ref{eq:fieldpressures}).  The scaling chosen in this figure renders the vertical axis dimensionless and has
the added benefit of making the scaled hard particle transverse pressure constant for better visualization.

\begin{figure}[t]
\includegraphics[width=0.5\textwidth]{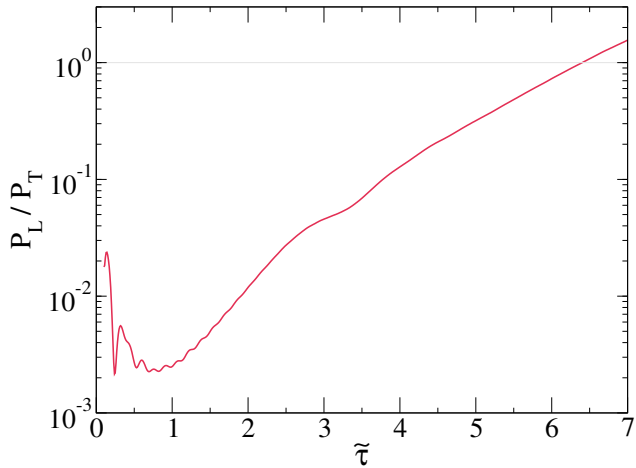}
\caption{(Color online)
Total longitudinal pressure over the total transverse pressure as a function of proper time.  The data were taken 
from the same set of runs as Fig.~\ref{fig:eDensity1}.}
\label{fig:pressure2}
\end{figure}

As can be seen 
from Fig.~\ref{fig:pressure1} the system is initially highly anisotropic with the transverse particle pressure dominating all other
contributions.  The $\tau$-scaled longitudinal particle pressure drops like $1/\tau^2$.
Note that at early times the field component
of the longitudinal pressure can become negative as evidenced by Fig.~\ref{fig:pressure1}.   This is consistent with the
finding of others \cite{Fukushima:2011nq} and is a result of coherent field modes. Without the unstable field growth, the
system would continue to become more and more anisotropic as time progresses and continue to experience positive and
negative pressure oscillations.  However, as Fig.~\ref{fig:pressure1} demonstrates, unstable field modes begin to generate a growing longitudinal field pressure that at late times dominates all other pressure
components.  

It should be noted, however, that by the time the longitudinal field pressure becomes of the same 
magnitude as the transverse particle pressure one already expects to see a significant amount of backreaction of the
hard particles on the unstable chromofields.  Physically this should result in a saturation of the field pressure growth due
to energy conservation.  In addition, the back reaction would serve to isotropize the particle sector.  
Such a physical saturation is, unfortunately, not describable in the hard-loop framework since
in this framework the hard particles act as an energy reservoir that can continue to pump energy into the soft
sector indefinitely.  Sans this caveat, we believe that this result shows evidence that the chromo-Weibel instability 
can restore isotropy on the fm/c time scale.

In Fig.~\ref{fig:pressure2} we show the total longitudinal pressure over the total transverse pressure 
(\ref{eq:totalpressure}) as a function of proper time.  The data were taken from the same set of runs as 
Fig.~\ref{fig:eDensity1}.  This plot condenses the information seen in the previous plot allowing one  to easily
see the point at which the plasma becomes isotropic in momentum space.  As can be seen from this figure this
occurs at approximately $\tilde\tau = 6.5$; however, the system continues to evolve beyond this
point with the total longitudinal pressure then exceeding the total transverse pressure.  This is most definitely
an artifact due to the lack of the back reaction of the hard particles on the chromofields.  Therefore,
we are only fully confident in the results we obtain at earlier times.

\begin{figure}[t]
\includegraphics[width=0.5\textwidth]{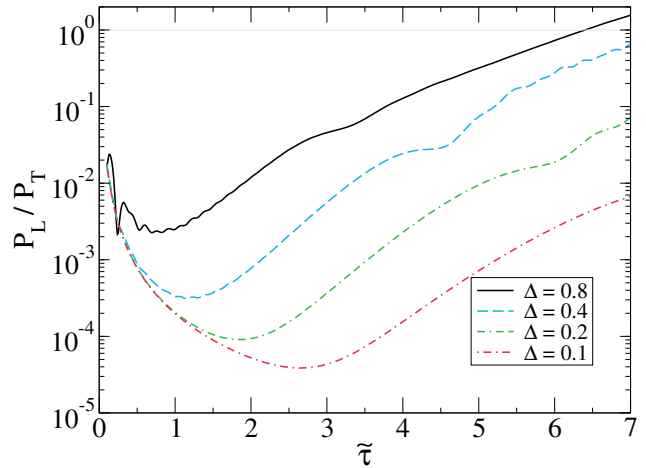}
\caption{(Color online)
Total longitudinal pressure over the total transverse pressure as a function of proper time for different initial current 
fluctuation magnitudes $\Delta \in \{0.1,0.2,0.4,0.8\}$.  The data were taken from the same runs as shown in 
Fig.~\ref{fig:eDensity2}.
}
\label{fig:pressure3}
\end{figure}

\begin{figure}[t]
\centering
\hspace{1.25cm}{\footnotesize (a)} \\
\includegraphics[width=0.47\textwidth]{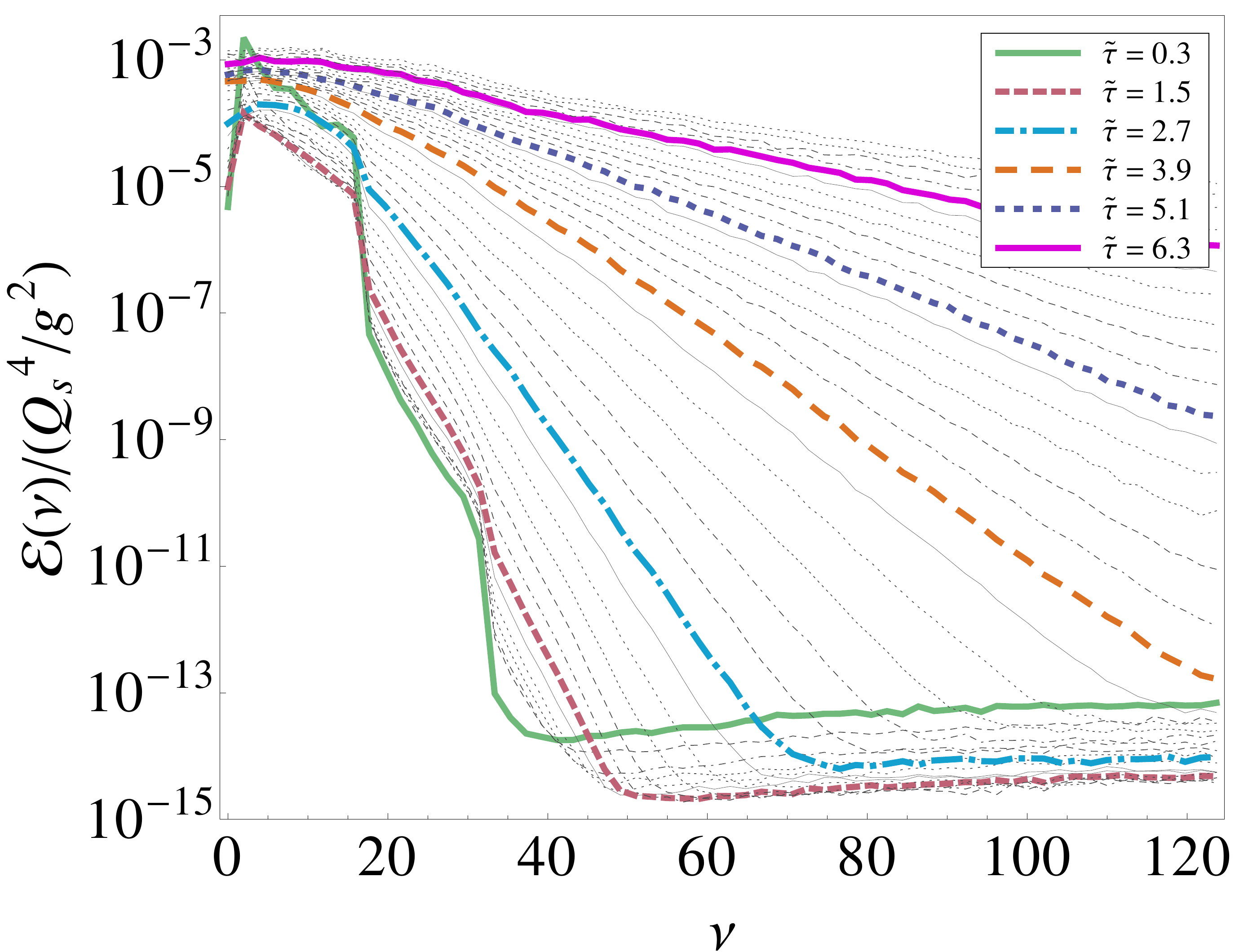}\\
\vspace{3mm}
\hspace{1.25cm}{\footnotesize (b)}
\includegraphics[width=0.47\textwidth]{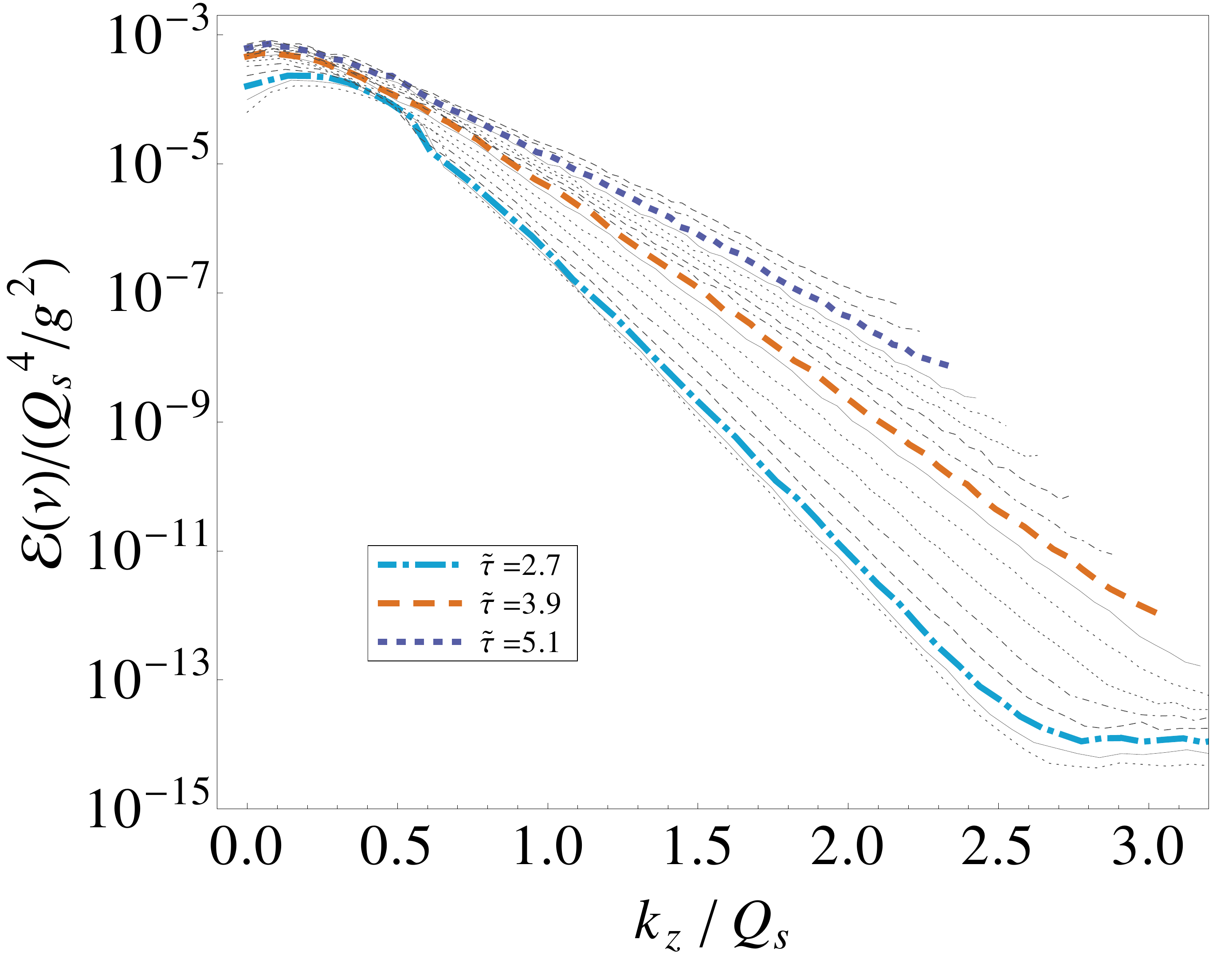}
\caption{(Color online)
The longitudinal energy spectra at various proper times as a function of (a) $\nu$ and (b) $k_z = \nu/\tau$.  Data 
taken from the averaged runs shown in Fig.~{\ref{fig:eDensity1}}.
}
\label{fig:spectra1}
\end{figure}

\begin{figure}[t]
\includegraphics[width=0.45\textwidth]{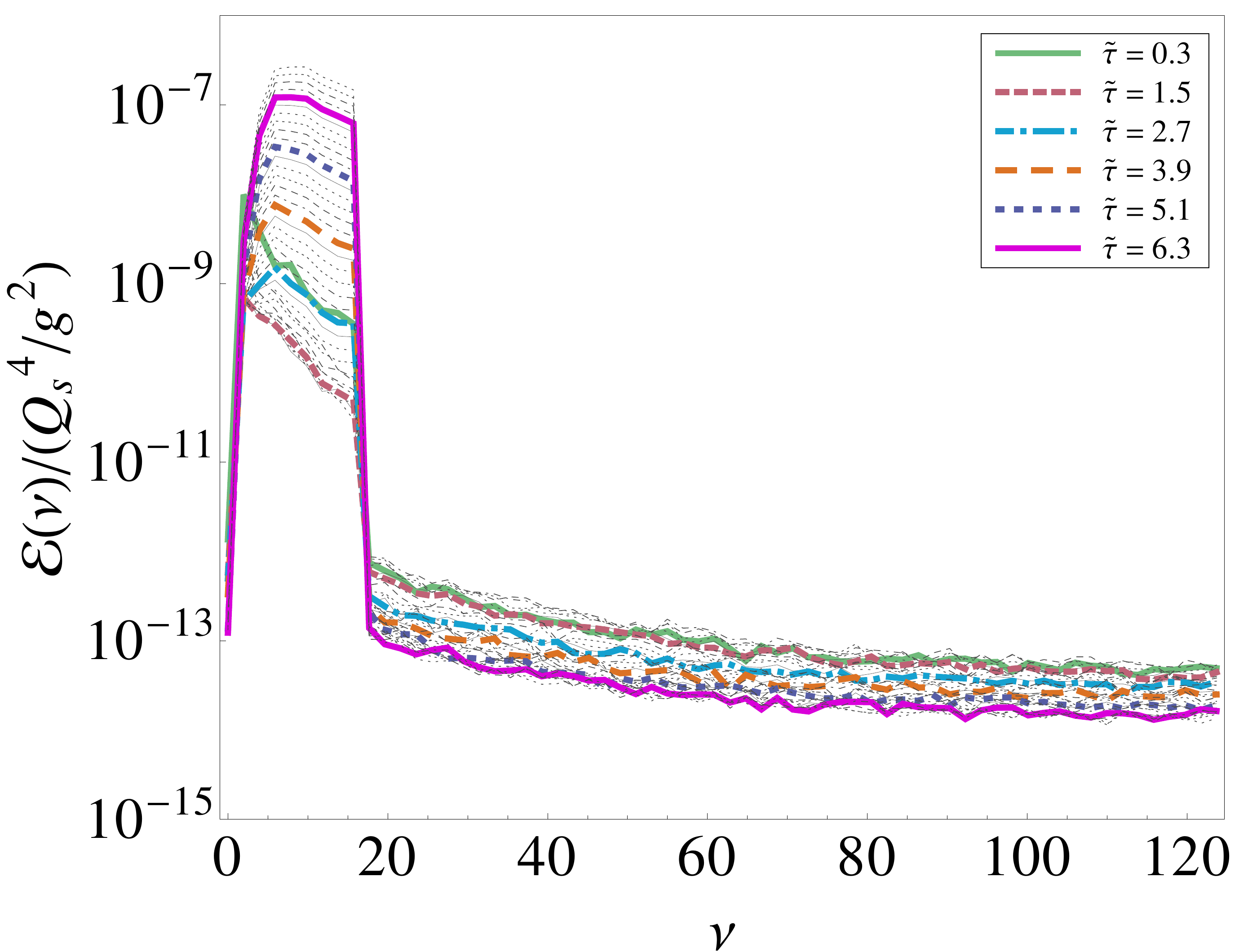}
\caption{(Color online)
The longitudinal energy spectra at various proper times as a function of $\nu$
for Abelian runs. For this figure the spectra from 40 runs were averaged.
}
\label{fig:spectra-Abelian}
\end{figure}

In Fig.~\ref{fig:pressure3} we show the total longitudinal pressure over the total transverse pressure 
(\ref{eq:totalpressure}) as a function of proper time for different initial current fluctuation magnitudes 
$\Delta \in \{0.1,0.2,0.4,0.8\}$.  The data were taken from the same runs as shown in Fig.~\ref{fig:eDensity2}.
The purpose of this figure is to show that the variable which has the biggest effect on the isotropization time
is our assumed magnitude of the initial current fluctuations, $\Delta$.  
{
From this figure we see that the isotropization time
scale depends roughly logarithmically on $\Delta$. In the limit of
parametrically small $\Delta$, where the evolution is dominated by
the Abelian behavior, one can infer from the analytical results
of Ref.~\cite{Romatschke:2006wg} that the square root of the
apparent isotropization time depends linearly on $\log\Delta^{-1}$
(which would lead to the estimate of $(\log g^{-1})^2$ for the
parametric dependence of isotropization time on $g$ in the
limit of weak coupling).
}


\subsection{Energy spectra}

In Fig.~\ref{fig:spectra1} we show the run-averaged longitudinal energy spectra obtained via (\ref{eq:longenergyspec}) 
at different proper times as a function of (a) the longitudinal wavenumber $\nu$ and (b) the longitudinal momentum 
$k_z = \nu/\tau$.  The data for both plots were taken from the same set of runs as Fig.~\ref{fig:eDensity1}.
In both figures the vertical axis is logarithmic while the horizontal axis is linear.  From Fig.~\ref{fig:spectra1} (a) 
we see the rapid emergence of an exponential distribution of longitudinal energy.  The exponential spectra persist
during the entire evolution.  In Fig.~\ref{fig:spectra1} (b) we show the spectra as function of the physical 
momentum so that one can now see the effect of the red-shifting of the longitudinal momentum with time.  In addition,
from this figure we can easily determine a kind of effective longitudinal temperature which can be extracted from the
slopes of the curves.  Below we will define a fit function and extract the longitudinal temperature as a function of
proper time.

Note that the emergence of this exponential spectrum is not solely due to the widening unstable
mode band.  Instead having nonlinear mode-mode coupling is vitally important in order to populate high momentum
modes which are rapidly becoming unstable as time progresses.  In order to illustrate this point in 
Fig.~\ref{fig:spectra-Abelian} we show the corresponding spectra from Abelian runs. The lattice size for these
Abelian runs were exactly the same as for the corresponding non-Abelian run shown in Fig.~\ref{fig:spectra1}; however,
we chose a smaller value of $\Delta$ in order to eliminate the possibility of artificial nonlinearities due to the fact that
we are simulating compact $U(1)$.  As we can see from this figure, only modes present in the initial conditions are
amplified in the Abelian case, hence demonstrating that the emergence of an exponential longitudinal energy spectrum 
is intrinsically non-Abelian (nonlinear).

At first sight our exponential distribution of longitudinal energy seems to be different than the result obtained 
by Fukushima and Gelis who saw the emergence of a power-law spectrum in Yang-Mills solutions in an expanding
QGP \cite{Fukushima:2011nq}; however, we note, importantly, that they saw the emergence
of a power-law longitudinal energy spectrum only at extremely late times corresponding to $\tilde\tau \gtrsim 150$.
At early times, their spectra also appear consistent with an exponential distribution of longitudinal 
energy.  Since we do not include the back reaction, we are unable to comment on the asymptotic behavior of the 
spectra since we currently see no evidence of soft-scale saturation of the unstable mode growth.  In addition,
power law scaling usually emerges in the infrared and, in that sense, we are limited due to small lattices. 

\begin{figure*}[t]
\includegraphics[width=0.99\textwidth]{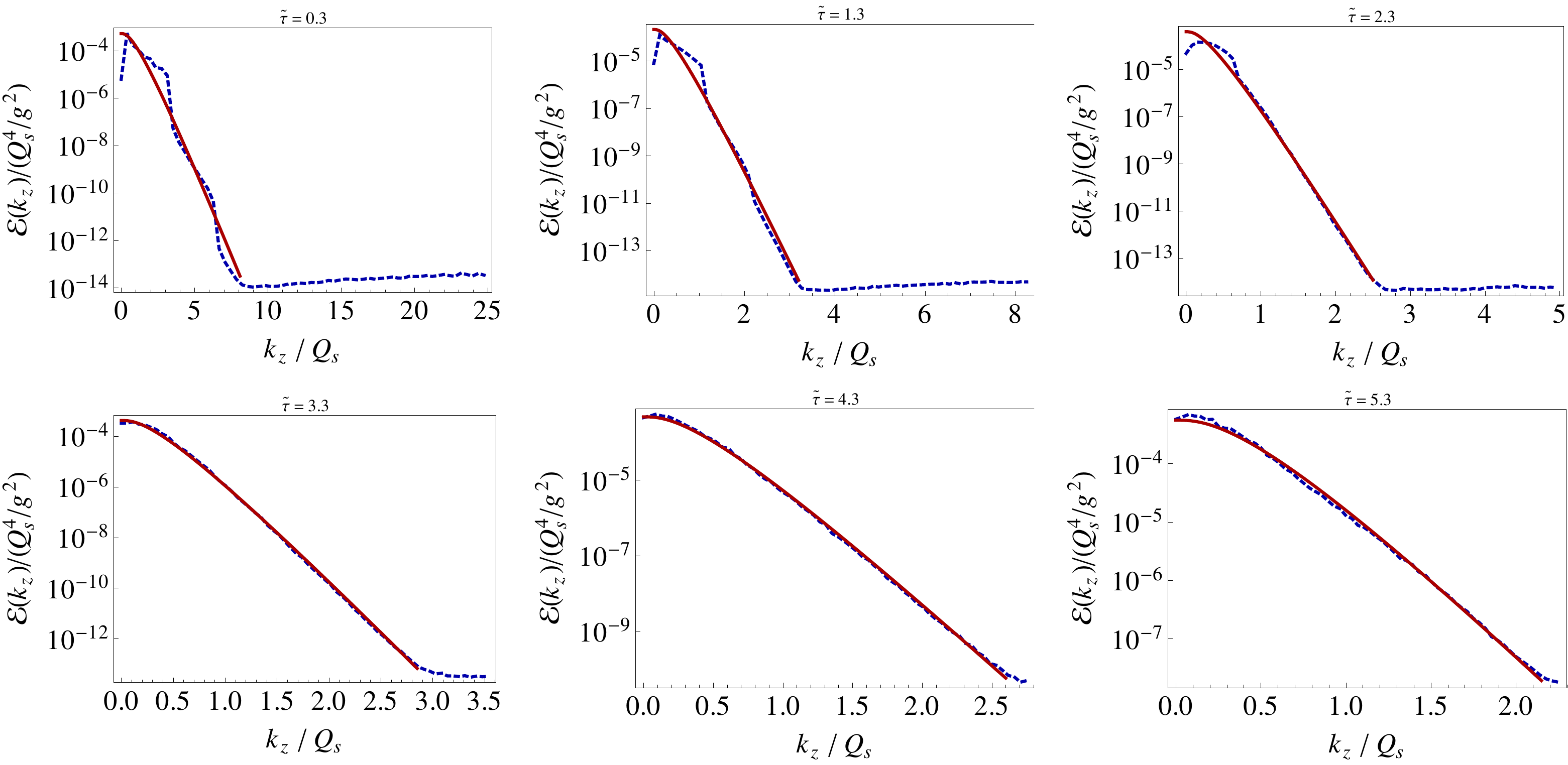}
\caption{(Color online)
Comparison of the longitudinal spectra data from Fig.~\ref{fig:spectra1} with fits using the fit function 
(\ref{eq:specfitfunc}) at six different proper times.
}
\label{fig:spectra2}
\end{figure*}

\begin{figure}[t]
\includegraphics[width=0.47\textwidth]{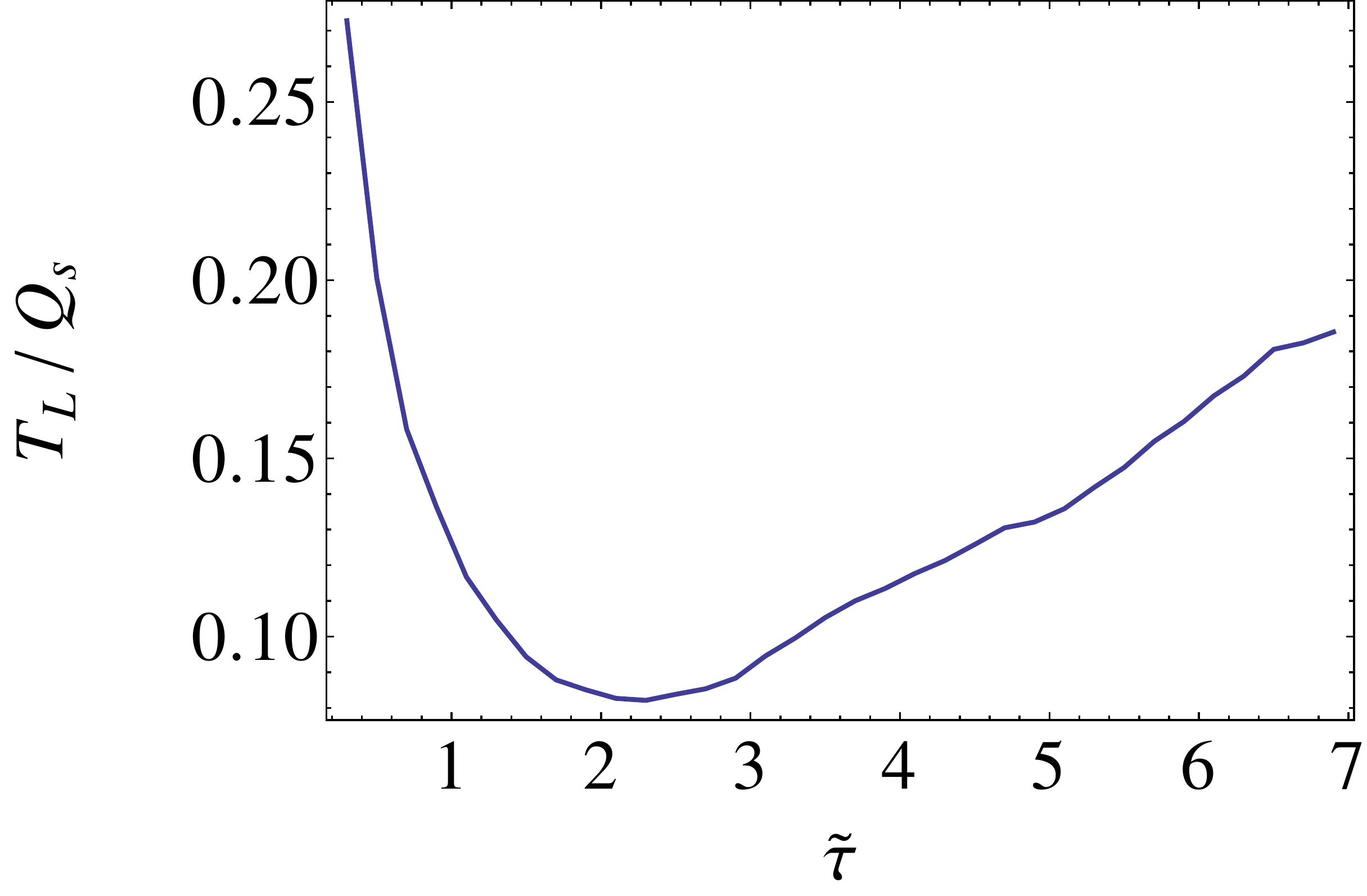}
\caption{(Color online)
The time dependent longitudinal temperature extracted from the data contained in Fig.~\ref{fig:spectra1} using
the fit function (\ref{eq:specfitfunc}).
}
\label{fig:spectra3}
\end{figure}

In Fig.~\ref{fig:spectra2} we show fits to spectra shown in Fig.~\ref{fig:spectra1} (b) at several different proper
times.  For the fit function we assumed that
the spectra corresponded to the energy density obtained from a massless Boltzmann distribution that has been 
integrated over transverse momenta
\bea
{\cal E} &\propto& \int d k_z d^2 k_T \, \sqrt{k_T^2 + k_z^2} \, \exp\left(-\sqrt{k_T^2 + k_z^2}/T\right) \, , 
\nonumber \\
&\propto& \int d k_z \left( k_z^2 + 2 |k_z| T + 2 T^2 \right) \exp\left(-|k_z|/T\right) \, .
\eea
The integrand in the above expression was taken as our fit function
\be
{\mathcal E}_{\rm fit}(k_z) = A \left( k_z^2 + 2 |k_z| T + 2 T^2 \right) \exp\left(-|k_z|/T\right) \, ,
\label{eq:specfitfunc}
\ee
where we have allowed for an overall multiplicative constant $A$.  At each proper time we fit the two parameters
$A$ and $T$; however, at early times we manually exclude regions of the spectra that are part of the ``noisy
plateau'' at high longitudinal momenta, e.g. $k_z \gtrsim 8 \, Q_s$ from the $\tilde\tau = 0.3$ 
panel shown in Fig.~\ref{fig:spectra2} are excluded from the fit data.

As can be seen from Fig.~\ref{fig:spectra2} we see evidence of a very rapid emergence of a Boltzmann 
longitudinal energy spectrum.  At $\tilde\tau = 0.3$ the fit is already working quite well with the bumps seen
in the spectra being nonlinear resonance ``copies'' of the initial theta-function-like distribution of longitudinal
energy.  By $\tilde\tau = 2.3$ virtually all information about the initial condition is gone and by 
$\tilde\tau = 3.3$ the system seems to exhibit an exceptional degree of longitudinal thermalization with all
information about the initial condition lost.  We only show six specific times in Fig.~\ref{fig:spectra2}, however,
at all simulation times $\tilde\tau \gtrsim 0.3$ the fits seem to work remarkably well.  We note,
importantly, that although the spectra shown in Fig.~\ref{fig:spectra2} are averaged over runs, one 
sees the emergence of such a Boltzmann spectrum on a run-by-run basis.  We have averaged over runs in 
order to remove statistical noise and improve the quality of the fits.

In Fig.~\ref{fig:spectra3} we show the extracted fit temperatures using (\ref{eq:specfitfunc}) as a function of 
proper-time.  We see from this figure that at early times the soft sector cools down due to longitudinal expansion,
but once the instability begins to grow, the soft sector begins to heat up.  We note in this context that the hard
particle distribution is highly anisotropic, making it hard to associate a temperature with.  The transverse temperature
given by $p_{\rm hard} \sim Q_s$ is a constant for longitudinal free streaming; however, one can associate a kind
of isotropic temperature by computing the fourth root of the energy density ${\cal E} = {\cal R}(\xi) 
{\cal E}_{\rm iso}(p_{\rm hard})$ \cite{Martinez:2010sc}.  One finds at late times ($\tau \gg \tau_{\rm iso}$) that ${\cal E} \sim \tau^{-1}$ so that $T_{\rm eff, hard}
\sim {\cal E}^{1/4} \sim \tau^{-1/4}$ which decreases less quickly than ideal hydrodynamical behavior for which one 
has $T \sim \tau^{-1/3}$.  Since the hard particles still dominate the energy density, the combined 
soft plus hard effective temperature still decreases in time.


\section{Conclusions}
\label{sec:conclusions}

In this paper we have studied the dynamics of the chromo-Weibel plasma instability in a 
longitudinally expanding plasma by numerically solving the full 3D+3V realtime evolution 
of the hard-loop equations of motion.  We utilized current fluctuations as the initial condition 
so that the initial fields were self-consistently triggered by the hard particles.  We had three 
important findings:  (1) there is no saturation of the chromo-Weibel instability at the 
``soft-scale'' on timescales relevant for heavy-ion collisions, (2) the dominant transverse 
chromomagnetic fields generate a rapidly 
growing longitudinal pressure that works to isotropize the system on timescales
relevant for heavy-ion collisions, and (3) in the process of evolution the longitudinal energy 
spectrum shows no signs of a power-law spectrum associated with Kolmogorov turbulence, but 
instead shows evidence for rapid longitudinal thermalization of the gauge fields.  

The finding that there is no soft-scale saturation of the plasma instability is important
since this means that on the time scales relevant for heavy ion collisions the back reaction 
of the hard degrees of freedom could be important.  This suggests that it will be of the 
upmost importance to make an in depth study of the dynamics of an unstable expanding
plasma using classical Yang-Mills and Boltzmann-Vlasov simulations.  However, care
will have to be taken to make sure that these simulations can properly describe the
soft collective modes of the system consistent with hard-loop dynamics in the high
temperature limit.  The fact that we do not witness soft-scale saturation of the 
chromo-Weibel instability is consistent with previous analyses of plasmas possessing
a fixed high-magnitude momentum-space anisotropy \cite{Bodeker:2007fw,Arnold:2007cg}.
In the case of HEL, the red shifting of the longitudinal momentum causes transverse
unstable modes to become more and more stable as a function of time, while purely 
longitudinal modes continue to grow.
We cannot rule out a very late time saturation on timescales
far beyond what we have studied; however, such large time scales are probably not
relevant to understanding thermalization of a QGP generated in heavy ion collisions.

Our second finding concerned plasma isotropization.  Extrapolating our
results to conditions expected for heavy-ion collisions at the LHC, we
found that for the assumed magnitude of current fluctuations, $\Delta
= 0.8$, isotropization within our framework occurs at $\sim$
6.5 fm/c.  Further extrapolating our numerical results to $\Delta =
1.6$ which is required in order to achieve occupation numbers
consistent with quantum fluctuations, one finds isotropization times
on the order of 5 fm/c.  {
However, it should be noted
  that we have not included the back-reaction of the hard particles on
  the soft background field.  It is likely that the back-reaction
  slows down the process of isotropization at late times and,
  therefore, the numbers quoted above should perhaps be taken as a
  lower bound on the time of {\em complete} isotropization.}

We note that although
the early indications from ideal hydrodynamics would imply that this time scale is much
too long, in recent years it has emerged that there is very little experimental constraint
on the degree of local momentum-space anisotropy in the quark gluon plasma.  In HEL
the precise time scale for isotropization depends on the choice of the amplitude of the initial
current fluctuations and as a consequence the amplitude of the soft gauge fields at 
early times.  We have chosen the magnitude of these fluctuations based on studies of
the breaking of boost invariance in the glasma by quantum fluctuations.  Of course, one can shorten
the isotropization time by increasing the magnitude of the initial fluctuations used; however,
within the hard-expanding loop framework one runs the risk of violating the assumption
that the energy of the system is dominated by the hard degrees of freedom.  Once again this
imposes a limit on what can be achieved through hard-loop simulations and
calls for more comprehensive methods to tackle the problem which can properly
include the back-reaction.

Our final finding concerned the induced spectrum of the unstable soft modes.   We
found a Boltzmann distribution of longitudinal energies instead of a power law distribution as
was found in static simulations.  Extrapolating to RHIC and LHC conditions, this result seems 
to imply that one can achieve longitudinal thermalization of the quark gluon plasma on time scales of 1 fm/c.  
Early color glass condensate simulations demonstrated that the initial gauge field configurations 
were transversally thermal \cite{Krasnitz:2001qu} and our results indicate that the system also 
quickly becomes thermal in the longitudinal direction.

The longitudinal thermalization we see is particular to non-Abelian gauge theories.  In general,
there are two effects occurring:  (1) mode amplification due to plasma instability and (2)
mode-mode coupling due to nonlinear interactions.  In an Abelian plasma only mode amplification
occurs and one does not see the emergence of a longitudinally thermalized spectrum.  One
needs the mode-mode coupling to spread the deposited energy across large ranges of 
momenta quickly.  In the non-Abelian case, we have checked different lattice sizes, lattice
spacing, etc.\ and the rapid emergence of a longitudinally thermalized spectrum seems to be 
quite robust.  

We note that our finding
of an exponential longitudinal energy spectrum is not in contradiction with the pure 
Yang-Mills simulations of  Ref.~\cite{Fukushima:2011nq} which found the emergence of 
a power-law spectrum, since the power-law spectrum observed therein only emerged at quite late
times, $\tau \gtrsim $~150 fm/c.  At early times Ref.~\cite{Fukushima:2011nq} also
found what appears to be an exponential distribution in the longitudinal energy spectrum.
We also note that usually one sees power law spectra energy in the infrared.  Due to
having to use many auxiliary fields we were limited to $40^2\times128$ lattices.  In the
future we plan runs on larger lattices in order to more carefully determine the infrared part
of the spectrum.

Our study, however, is not without caveats.  In order to have a tractable way to treat the
time-dependent hard particles, we approximated them as a longitudinally free streaming 
ensemble.  This is an extreme assumption that should be relaxed, if possible, in the future.
Some work along this direction has been started in Ref.~\cite{Florkowski:2012ax} where
the authors were able to derive an evolution equation for a stable uniform chromoelectric 
field in an arbitrary time-evolving anisotropic background.  It would be very interesting to 
see if the method employed in Ref.~\cite{Florkowski:2012ax} can be extended to the entire 
stable and unstable mode spectrum.
This caveat aside, it's interesting that even with such an extreme particle pressure anisotropy
being developed, the chromo-Weibel instability is able to isotropize the system on time
scales relevant for heavy-ion collisions.  

The second important caveat is that we did not include 
the effect of the back reaction of the hard particles on the unstable soft gauge fields.  Our results 
seem to indicate that the fields grow unabated until there will be a significant backreaction.  
Of course, as soon as the field amplitudes become large enough for any back reaction 
to occur, it is possible that this could reduce the anisotropy of the hard-particles and reduce 
the rate of growth of the unstable soft modes.

In the context of our numerical results, the observation of continued unstable mode
growth places an upper limit on the amount of time over which we can trust our hard-loop 
simulations; however, we find that assuming that the initial fraction of the energy carried 
by soft fields
is small compared to the hard scale there is a window of time over which we can reliably 
simulate the dynamics.  Our results indicate a very fast path to isotropization within this
window of reliability.  Addressing the question of the late time dynamics of the system is
not possible within this framework;\footnote{In this paper, we have concentrated on the
phenomenology of unstable modes in a longitudinally free streaming background.  For an 
in-depth analysis of the  path to isotropy in the asymptotically small coupling limit, including 
late time dynamics, we refer the reader to Refs.~\cite{Kurkela:2011ti,Kurkela:2011ub} where
parametric estimates have been made.} however, our study 
might serve as a benchmark for future simulations that include backreaction in an expanding 
plasma.

In the future one might use hard-loop simulations to study the early time dynamics of
the quark gluon plasma and the role unstable modes play.  One can address interesting
phenomenological questions such as measuring the shear viscosity due to plasma instabilities
and studying particle transport properties such as energy loss and momentum-space 
diffusion.  In addition, the momentum-space anisotropy dependence of many important 
heavy-ion collision observables such as jet energy loss, photon production, dilepton 
production, heavy quark energy loss, heavy quarkonium suppression etc. have been 
computed \cite{Romatschke:2003vc,Romatschke:2004au,Romatschke:2006bb,%
Mauricio:2007vz,Dumitru:2007rp,Dumitru:2007hy,Martinez:2008di,Dumitru:2009ni,%
Majumder:2009cf,Dumitru:2009fy,Burnier:2009yu,Philipsen:2009wg,Mannarelli:2009pd,%
Margotta:2011ta,Strickland:2011mw,Strickland:2011aa,Carrington:2011uj}.  It would,
therefore, be interesting to study the effect of our time-dependent evolution on these 
observables as a possible signature of the plasma instability in heavy ion collisions.  

We note that there are now many groups studying the thermalization, isotropization, and 
anisotropic signatures
of the quark gluon plasma in the strong coupling limit using the AdS/CFT correspondence
\cite{Chesler:2008hg,Janik:2008tc,Chesler:2009cy,Mateos:2011ix,Mateos:2011tv,Heller:2011ju,%
Heller:2012je,Heller:2012km,Wu:2011yd,Chesler:2011ds,Chernicoff:2012iq,Chernicoff:2012gu,%
Giataganas:2012zy,Rebhan:2011ke,Rebhan:2011vd,Rebhan:2012bw}.  
It would be interesting to
compare and contrast the predictions for experimental observables coming from the 
weakly-coupled and strongly-coupled frameworks. 

Finally, there have also been some 
recent studies that have suggested that there is an inverse particle number cascade 
leading to Bose-Einstein condensation of soft-gauge fields \cite{Blaizot:2011xf,%
Berges:2012us,Berges:2012ev}.  How a long-lived condensate can emerge in a non-Abelian
gauge is an open question.  Based on our results it is hard to judge whether this possibility
is borne out, since we do not directly obtain the particle number spectra but instead the
energy spectra.  Determining the nature of the low momentum number spectra is complicated
by gauge invariance issues; however, measurements of this spectra in fixed-anisotropy hard-loop 
simulations \cite{Arnold:2005qs,Strickland:2007fm,Ipp:2010uy} and pure Yang Mills with
high occupancy \cite{Berges:2012ev,Kurkela:2012hp,Schlichting:2012es} have so far shown no evidence 
of occupation numbers exceeding $f \sim 1/\alpha_s$ at late times.  That being said it would be interesting
to see if a time-evolving condensate, perhaps in the form of an overpopulated 
condensate of plasmons (chromoelectric oscillations), could play
a role in QGP thermalization and isotropization.

\begin{acknowledgments}
We would like to thank J. Berges, K. Fukushima, F.  Gelis,  A. Ipp, A. Kurkela,  G.D. Moore, and 
A. Mueller for useful discussions.  We thank the 
Vienna Scientific Cluster for providing computational resources under project 
no.\ 70061.  M.A. was supported by the Austrian Science Fund (FWF) grant no.~P19526, P22114, 
the START project Y435-N16, and the Austrian Marshall Plan Foundation.  M.S. was supported by 
NSF grant no.~PHY-1068765 and the Helmholtz International Center for FAIR LOEWE program.
\end{acknowledgments}

\appendix

\section{Lattice equations of motion}
\label{app:eom-lattice}

In this appendix we introduce the dimensionless lattice variables we use in simulating the dynamics
of the soft color fields.  We then explicitly write the discretized equations of motion and initial 
conditions used in the main body of the paper.  In this paper we consider the non-Abelian SU(2) group; 
however, the equations below are independent of the gauge group considered.

\subsection{Lattice variables}

We begin by defining dimensionless lattice variables which will be used in the simulation.
We introduce three lattice spacings:  $a$ which is the dimensionful transverse spatial lattice spacing,
$\epsilon$ which is the dimensionful temporal lattice spacing, and $a_\eta$ which is the 
dimensionless lattice spacing in the $\eta$ direction.
We rescale space and time
\bea
&& {\hat x} = x/a \, , \quad {\hat y} = y/a \, , \quad {\hat \eta} = \eta \, , \nonumber \\
&& {\hat \tau} = \tau/a \, , \quad {\hat \epsilon} = \epsilon/a \, ,
\label{eq:spatial-rescalings}
\eea
With these definitions we can rescale the field variables, conjugate momenta,
and currents and introduce lattice variables with ``hats''
\bea
{\hat A}^i = g a A^i \, , 
\quad &&
{\hat A}_\eta = g A_\eta \, ,
\nonumber \\
{\hat \Pi}_i = g a \Pi_i \, ,
\quad &&
{\hat \Pi}^\eta = g a^2 \Pi^\eta \, ,
\label{eq:field-rescalings}
\eea
and
\bea
{\hat{\overline{\mathcal W}}} = a \overline{\mathcal W}  \, , 
\quad &&
{\hat j}^\tau = a^3 j^\tau \, ,
\nonumber \\
{\hat j}^i = a^3 j^i\, ,
\quad &&
{\hat j}^\eta = a^4 j^\eta \, .
\label{eq:current-rescalings}
\eea
Finally, we rescale the isotropic Debye mass via $\hat{m}_D = a \,  m_D$.
Performing this transformation on the Hamiltonian density we find
\be
{\mathcal H} = 
\frac{\hat\tau}{g^2 a^3} \tr\left[\frac{1}{\hat\tau^2}{\hat F}_{\eta i}^2 +
\frac{1}{\hat\tau^2} \hat\Pi_{i}^2 +
{\hat F}_{xy}^2+ \left(\hat\Pi^{\eta}\right)^2 \right] .
\ee
In the following subsections we will drop the ``hats'' on symbols. From
this point on in this appendix, all variables can be assumed to be dimensionless 
lattice variables.

\subsection{Plaquettes and Staples}

We can translate the continuum equations of motion into gauge-invariant
lattice equations of motion by using standard plaquette and staple operators.
For the transverse coordinate-rapidity plaquettes we have
\be
(F_{k\eta})^a =  \frac{i N_c}{a_\eta} \tr\left[ \tau^a U_{\Box,k\eta} \right] \, ,
\label{fketaplaquette}
\ee
where $k \in \{x,y\}$, $a$ is a color algebra index, $a \in \{1, \cdots , N_c^2-1 \}$,
and $U_{\Box,\mu\nu}(x)=U_{\mu}(x)U_{\nu}(x+\mu)U_{\mu}^\dagger(x+\nu) 
U_{\nu}^\dagger(x)$ is a standard lattice plaquette variable with $U_\mu$ being 
a parallel transporter in $\mu$ direction 
\bea
U_i &=& {\rm exp}(-i A^i) \, , \\
U_\eta &=& {\rm exp}(i a_\eta A_\eta) \, .
\eea

Products like $F_{\mu \nu}^2$ which appear in the energy density can
also be expressed in terms of plaquette variables.  For this application we need two
different combinations, $F_{\eta i}^2$ and $F_{xy}^2$.  These are
\bea
\tr \, F_{\eta i}^2 &=& \frac{2}{a_\eta^2} \left( 1 - \frac{1}{N_c} \tr[{\rm Re}\, U_{\Box,\eta i}] \right) ,
\label{plaquetteFetai}
\\
\tr \, F_{xy}^2 &=& 2 \left( 1 - \frac{1}{N_c} \tr[{\rm Re}\, U_{\Box,xy}] \right) .
\label{plaquetteFxy}
\eea

Finally, we can rewrite the necessary covariant derivatives acting on the field strength tensor as 
\bea
(D_j F_{jk})^a &=& i N_c \tr \biggl[ \tau^a U_k(\tau,x) \!\sum_{|j|\neq k} \! S_{kj}^{\dagger}(\tau,x) \biggr] \! ,
\nonumber \\ \\ 
(D_\eta F_{\eta j})^a &=& \frac{i N_c}{a_\eta^2} \tr \biggl[ \tau^a U_j(\tau,x) \!\sum_{|\eta|\neq j} \! S_{j\eta}^{\dagger}(\tau,x) \biggr] \! ,
\nonumber \\ \\
(D_j F_{j\eta})^a &=& \frac{i N_c}{a_\eta} \tr \biggl[ \tau^a U_\eta(\tau,x) \!\sum_{|j|\neq \eta} \! S_{\eta j}^{\dagger}(\tau,x) \biggr] \! ,
\nonumber \\ 
\label{covdervs}
\eea
where $S$ is the gauge link staple
\be
S_{\mu
\nu}^{\dagger}(\tau,x)=U_\nu(\tau,x+\mu)U_\mu^{\dagger}(\tau,x+\nu)
U_\nu^{\dagger}(\tau,x) \, .
\ee
Note that the sums in (\ref{covdervs}) run over both positive and negative directions.

\subsection{Transformation of the $\overline{\mathcal W}$ fields to a compact domain}

In order to better describe $\overline{\mathcal W}$ in the $\bar y$ (shifted rapidity) direction we introduce a 
velocity-like variable $u$, $-1 < u < 1$, defined by
\be
\bar y \equiv {\rm atanh }(u)\,,\quad d\bar y = \frac{1}{1 - u^2} du.
\ee
This has the effect of giving more lattice points around $\bar y = 0$, 
where the $\overline{\mathcal W}$ functions are rapidly varying.

Using $\sinh^2(\bar y) = {u^2}/({1 - u^2})$ and $\cosh^2(\bar y) = 1/({1 - u^2})$ we
can rewrite (\ref{DVW3+1v2}) as
\bea
\label{DVW3+1v3}
&& \partial_\tau \overline{\mathcal W}(\tau,{\bf x},\eta;\phi,u) = 
-\sqrt{1-u^2} \, v^{i} D_{i} \overline{\mathcal W} 
\nonumber \\
&&
- \frac{u}{\tau} \left( D_\eta\overline{\mathcal W} 
- (1-u^2) \partial_u\overline{\mathcal W} \right) 
\nonumber \\ && 
+ \frac{1}{\bar f(\tau,\tau_{\rm iso},u)} 
\biggl[ \frac{1}{\tau}v^{i} \Pi_{i} -
\nonumber \\
&&
\frac{\tau^2 }{ \tau_{\rm iso}^2} \frac{u }{ \sqrt{1-u^2}} \Pi^\eta 
+ \frac{u}{\tau} \left(1- \frac{\tau^2}{\tau_{\rm iso}^2} \right)
v^{i} F_{{i}\eta} \biggr]
\, ,
\eea
where
\be
\bar f(\tau,\tau_{\rm iso},u) = \left(1+\frac{\tau^2}{\tau_{\rm iso}^2} \frac{u^2 }{ 1 - u^2} \right)^2 \, .
\ee
The currents are then given by
\bea
j^\tau &=& -\frac{m_D^2}{2} \int_0^{2 \pi} \frac{d\phi}{2\pi} \int_{-1}^1 du \, (1 - u^2)^{-\frac{3}{2}}
 \, \overline{\mathcal W}(\tau,{\bf x},\eta;\phi,u)  \,, \nonumber \\ 
\\
j^i &=& -\frac{m_D^2}{2} \int_0^{2 \pi} \frac{d\phi}{2\pi} \int_{-1}^1 du \, v^i \, (1 - u^2)^{-1}
 \, \overline{\mathcal W}(\tau,{\bf x},\eta;\phi,u)  \,, \nonumber \\
\\ 
j^\eta &=& -\frac{m_D^2}{2\tau} \int_0^{2 \pi} \frac{d\phi}{2\pi} \int_{-1}^1 du \, u \, (1 - u^2)^{-\frac{3}{2}}
 \, \overline{\mathcal W}(\tau,{\bf x},\eta;\phi,u)  \, , \nonumber \\
\eea
where, as usual, $v^i = (\cos\phi, \sin\phi)$ with $i \in \{x,y\}$.

\subsection{Lattice equations of motion}

We will express the equations of motion in terms of gauge links $U$ and chromoelectric fields $\Pi$.
Both $U$'s and $\Pi$'s live on links (between sites) so all of their spatial arguments have an implicit $+1/2$ shift.
In some cases this $1/2$ is made explicit for maximum clarity.  Temporally $\Pi$'s also live between sites.  The 
link variables $U$, however, temporally live on sites.  The ${\mathcal W}$'s and $j$'s live on sites both spatially 
and temporally.  We use a lattice with $N_\perp$ sites in the $x$ and $y$ directions and $N_\eta$ sites 
in the $\eta$ direction.  The fields are assumed to be periodic in all directions.  We use a leapfrog algorithm
in which the conjugate momenta are updated first using fixed links/currents
and then the link variables and ${\cal W}$-fields are evolved using the updated conjugate momenta 
\cite{Krasnitz:1995xi,Ambjorn:1995xm,Moore:1996wn,Grigoriev:1989je,Ambjorn:1990pu}.


\begin{widetext}
The resulting Yang-Mills update equations are
\bea
\Pi_{i}(\tau+\frac{\epsilon}{2},{\bf x},\eta)&=&\Pi_{i}(\tau-\frac{\epsilon}{2},{\bf x},\eta)
+ \tau \epsilon \, \biggl( j^i_{\rm avg}(\tau,{\bf x},\eta) 
+ D_j F_{ji}(\tau,{\bf x},\eta) + \frac{1}{\tau^2} D_\eta F_{\eta i}(\tau,{\bf x},\eta) \biggr) , \\
\Pi^{\eta}(\tau+\frac{\epsilon}{2},{\bf x},\eta)&=&\Pi^{\eta}(\tau-\frac{\epsilon}{2},{\bf x},\eta) - 
\frac{\epsilon}{\tau} \biggl( \tau^2 j^\eta_{\rm avg}(\tau,{\bf x},\eta) + D_i F_{i\eta}(\tau,{\bf x},\eta) \biggr) , \\
U_{i}(\tau+\epsilon,{\bf x},\eta)&=&\exp{\!\left(-\, i \, \epsilon\ \tau^{-1} \
\Pi_{i}(\tau+\frac{\epsilon}{2},{\bf x},\eta)\right)} \, U_{i}(\tau,{\bf x},\eta) \, , \\
U_{\eta}(\tau+\epsilon,{\bf x},\eta)&=&\exp{\!\left(+\, i \, \epsilon \, \tau \, a_\eta \,
\Pi^{\eta}(\tau+\frac{\epsilon}{2},{\bf x},\eta)\right)} \, U_{\eta}(\tau,{\bf x},\eta) \, ,
\eea
where
\bea
j^i_{\rm avg}(\tau,{\bf x},\eta) &\equiv& \frac{1}{2} \biggl[ j^i(\tau,{\bf x},\eta) 
+ U_i^\dagger(\tau,{\bf x},\eta) j^i(\tau,{\bf x}+\hat{\bf e}_i,\eta) U_i(\tau,{\bf x},\eta) \biggr] ,
\\
j^\eta_{\rm avg}(\tau,{\bf x},\eta) &\equiv& \frac{1}{2} \biggl[ j^\eta(\tau,{\bf x},\eta)  
+ U_\eta^\dagger(\tau,{\bf x},\eta) j^\eta(\tau,{\bf x},\eta+1) U_\eta(\tau,{\bf x},\eta) \biggr] . 
\eea
To discretize the $\overline{\mathcal W}$ fields we use a rectangular lattice in $\phi$-$u$ space of size
$N_\phi \times N_u$ and
\bea
\phi_n &=& 2 \pi n / N_\phi \, ,\\
u_m &=& -1  + (2m +1)/N_u \, ,
\eea
where $n \in \{0,\cdots,N_\phi-1\}$ and $m \in \{0,\cdots,N_u-1\}$.

\end{widetext}

The update equations for the $\overline{\mathcal W}$ fields then take the form
\bea
\label{Wupdate}
&& \overline{\mathcal W}(\tau+\epsilon,{\bf x},\eta;\phi,u) =
\nonumber \\
&&  \hspace{5mm} \overline{\mathcal W}(\tau-\epsilon,{\bf x},\eta;\phi,u) + 2 \epsilon \biggl\{
-\sqrt{1-u^2} \, v^{i} D_i^S \overline{\mathcal W} 
\nonumber \\
&&  \hspace{5mm}
- \frac{u}{\tau} \left( D_\eta^S\overline{\mathcal W} 
- (1-u^2) \partial_u^S\overline{\mathcal W} \right) 
\nonumber \\ 
&& \hspace{5mm}  
+ \frac{1}{\bar f(\tau,\tau_{\rm iso},u)} 
\biggl[ \frac{1}{ \tau}v^{i} \Pi_i^{\rm avg} -\frac{\tau^2 }{ \tau_{\rm iso}^2} \frac{u }{ \sqrt{1-u^2}} \Pi^\eta_{\rm avg}
\nonumber \\ &&  \hspace{3cm}
+ \frac{u}{\tau} \left(1- \frac{\tau^2}{\tau_{\rm iso}^2} \right)
v^{i} F_{{i}\eta} \biggr] \biggr\}
\, ,
\eea
where $F_{{i}\eta}$ is computed using plaquettes via Eq.~(\ref{fketaplaquette}),
$D_i^S$ and $D_\eta^S$ are symmetric covariant derivatives in the transverse and rapidity
directions, respectively
\bea
D^S_\eta \varphi(\eta) &\equiv&
\frac{1}{2a_\eta} \Biggl( U^\dagger_\eta(\eta) \varphi(\eta+1) U_\eta(\eta) 
\nonumber \\ && \hspace{5mm}
- U_\eta(\eta-1) \varphi(\eta-1) U^\dagger_\eta(\eta-1) \Biggr) \, ,
\nonumber \\ 
\eea
\bea
D^S_i \varphi(x_i) &\equiv&
\frac{1}{2} \Biggl( U^\dagger_i(x_i) \varphi(x_i+1) U_i(x_i) 
\nonumber \\ && \hspace{5mm}
- U_i(x_i-1) \varphi(x_i-1) U^\dagger_i(x_i-1) \Biggr) \, ,
\nonumber \\ 
\eea
and $\partial_u^S$ is a symmetric derivative in $u$ space
\be
\partial^S \varphi(u_m) \equiv
\frac{\varphi(u_{m+1}) - \varphi(u_{m-1}) }{2 \Delta u} \, ,
\ee
where $\Delta u = 2/N_u$.
The averaged conjugate momenta, $\Pi_i^{\rm avg}$ and $\Pi^\eta_{\rm avg}$, appearing in
(\ref{Wupdate}) are averaged both spatially and temporally
\begin{widetext}
\bea
\Pi_i^{\rm avg}(\tau,{\bf x},\eta) \! &=& \! \frac{1}{16}
\sum_{\sigma_x=\pm1} \sum_{\sigma_y=\pm1}\sum_{\sigma_\eta=\pm1}\sum_{\sigma_\tau=\pm1} \!\!
{\cal P_T}\Pi_i\left(\tau + \frac{\sigma_\tau}{2},x + \frac{\sigma_x}{2},y + \frac{\sigma_y}{2},\eta + \frac{\sigma_\eta}{2}\right) \!\! ,
\\
\Pi^\eta_{\rm avg}(\tau,{\bf x},\eta) \! &=& \! \frac{1}{16}
\sum_{\sigma_x=\pm1} \sum_{\sigma_y=\pm1}\sum_{\sigma_\eta=\pm1}\sum_{\sigma_\tau=\pm1} \!\!
{\cal P_T}\Pi^\eta\left(\tau + \frac{\sigma_\tau}{2},x + \frac{\sigma_x}{2},y + \frac{\sigma_y}{2},\eta + \frac{\sigma_\eta}{2}\right) \!\! ,
\eea
\end{widetext}
where we have indicated explicitly the fact that the $\Pi$'s live on links (halfway between sites) for clarity
and ${\cal P_T}$ stands for the parallel transporter necessary to bring the conjugate momenta to the 
same site.

The  currents are computed from the $\overline{\mathcal W}$ fields via
\bea
j^\tau(\tau,{\bf x},\eta) &=& -\frac{m_D^2}{N_\phi N_u} \sum_{n,m} \, (1 - u^2)^{-\frac{3}{2}}
 \, \overline{\mathcal W}(\tau,{\bf x},\eta;\phi,u)  \,, \nonumber \\ \\
j^i(\tau,{\bf x},\eta) &=& -\frac{m_D^2}{N_\phi N_u} \sum_{n,m} \, v^i \, (1 - u^2)^{-1}
 \, \overline{\mathcal W}(\tau,{\bf x},\eta;\phi,u)  \,, \nonumber \\ \\
j^\eta(\tau,{\bf x},\eta) &=& -\frac{m_D^2}{\tau N_\phi N_u} \sum_{n,m} \, u \, (1 - u^2)^{-\frac{3}{2}}
 \, \overline{\mathcal W}(\tau,{\bf x},\eta;\phi,u)  \, , \nonumber \\
\eea
and we monitor Gauss' Law by periodically checking 
\bea
&& \tr\Biggl[\frac{1}{N_\perp^2 N_\eta} \sum_{{\bf x},\eta}  
\tau j^\tau(\tau,{\bf x},\eta) + D^S_\eta \Pi^\eta_{\rm avg}(\tau,{\bf x},\eta) 
\nonumber \\ && \hspace{3.5cm}
- D_i^{S} \Pi_i^{\rm avg}(\tau,{\bf x},\eta)
\Biggr]^2 \, .
\eea
We compute the discretized transverse and longitudinal contributions to the 
field energy density ${\mathcal E}$ via
\bea\label{eq:pressure_fields}
{\mathcal E}_T &=& \frac{1}{N_\perp^2 N_\eta} \sum_{{\bf x},\eta} \tr\left[ \tau^{-2} F_{\eta i}^2 +
\tau^{-2}\Pi_{i}^2  \right] ,
\\
{\mathcal E}_L &=& \frac{1}{N_\perp^2 N_\eta} \sum_{{\bf x},\eta} \tr\left[ F_{xy}^2+ \left(\Pi^{\eta}\right)^2 \right] ,
\eea
where $\tr F_{\eta i}^2$ and $\tr F_{xy}^2$ are computed using Eqs.~(\ref{plaquetteFetai}) and (\ref{plaquetteFxy}).

\section{Choice of lattice parameters}
\label{app:eom-lattice-params}

In this appendix we detail the constraints which should be obeyed in order for our simulations to properly describe the 
soft gauge field dynamics.  Since the soft scale is time dependent, we have to choose parameters which allow
for a faithful representation of the infrared and ultraviolet physics during the entirety of the simulation.

The physical (dimensionful) parameter $m_D^2$ is the Debye mass at time $\tau_{\rm iso}$.  
In terms of the gluon liberation factor $c$ which is ${\cal O}(1)$ ($\simeq 1.1$ according to Lappi 
\cite{Lappi:2007ku}, $c=2\ln2\approx 1.386$ according to Kovchegov \cite{Kovchegov:2000hz}) one has 
\be
m_D^2 \tau_{\rm iso}\tau_0\approx 0.93 \, c \, (Q_s \tau_0) \, .
\ee
In the text we use $Q_s \tau_0=1$ and $c = 2\ln2$ from Kovchegov \cite{Kovchegov:2000hz}.  
This gives $m_D^2 \tau_{\rm iso}\tau_0=1.285.$   
For large anisotropy one finds 
\be
m_\infty^2(\tau)\simeq \frac{\pi}4 m_D^2\tau_{\rm iso}/\tau, 
\ee
which can be taken as the typical (time-dependent) soft momentum scale.  
With our choice of $c=2\ln 2$,
we have $m_\infty(\tau)\approx 1.0 \, (\tau_0 \tau)^{-1/2}$. For RHIC energies one has $\tau_0^{-1}=Q_s\sim 1.4$ 
GeV and at current LHC energies one has $Q_s\sim 2$ GeV. At RHIC and LHC energies $\tau_0=Q_s^{-1}$ 
corresponds to 0.14 fm/c and 0.1 fm/c, respectively.

On a lattice with periodic boundary conditions, the size of the lattice
determines the infrared cutoff in full wavelengths and the lattice
spacing determines the ultraviolet cutoff via the smallest half-wavelength.
In our expanding system, the transverse UV cutoff is constant in time
and given by $\pi/a$, whereas the soft momentum scale is decreasing in
time. It is therefore sufficient to ensure
\be 
k_{\rm max} = \frac{\pi}a \gg m_\infty(\tau_0) \, ,
\ee
so that we should demand $a \lesssim 1 \, \tau_0$.

In longitudinal direction, the effective UV cutoff is decreasing in time according to $\pi/(\tau a_\eta)$. We may 
choose for example
\be
\frac{\tau_{\rm max}}2\, a_\eta \sim a \, ,
\ee
to have comparable transverse and longitudinal UV cutoffs in an average sense.
More importantly, the maximal longitudinal wave number of unstable modes 
increases in time, so $\nu_{\rm max}$ should be a large number,
\be\label{eq:long_uv_cutoff}
\nu_{\rm max} = \frac{\pi}{a_\eta} \gg 30 \, .
\ee

Since the hard-expanding-loop framework is designed to treat the soft sector of the dynamics, it is
somewhat more important to properly treat the infrared scale.
In the longitudinal direction $\nu_{\rm min}=2\pi/(N_\eta a_\eta)$ should be made as low as possible. 
There are no important unstable modes with $\nu$ much smaller than 5, but $\nu_{\rm min}$ also sets 
the spacing between mode numbers. We should therefore aim at
\be
\nu_{\rm min}= \frac{2\pi}{N_\eta a_\eta} \ll 5 \, .
\ee
In the transverse direction, the semianalytic results \cite{Romatschke:2006wg,Rebhan:2009ku} suggest that 
we should have
\be
k_{\rm min}= \frac{2\pi}{N_\perp a} \ll 0.2\, \tau_0^{-1} \, .
\ee

As our canonical set of parameters in the results section we use $N_T = 40$, $N_\eta = 128$, $a_\eta = 0.025$,
$a = Q_s^{-1}$, and $\tau_0 = Q_s^{-1}$.  Checking the transverse infrared cutoff one finds $k_{\rm min} 
= 0.157 \, Q_s < 0.2 \, Q_s$ as required.    Checking the transverse ultraviolet cutoff one finds $k_{\rm max}
= \pi Q_s > 1.005\,Q_s$ as required.  Checking the longitudinal infrared cutoff one finds $\nu_{\rm min} = 1.96 < 5$
as required.  Finally, checking the longitudinal ultraviolet cutoff one finds $\nu_{\rm max} = 125.7
> 30$.

\section{Initial Conditions}
\label{app:eom-lattice-initialcondition}

In this appendix we collect details of the initial conditions used in the simulations and some information 
about lattice initial conditions in general.

\subsection{Gaussian random variables}

We now discuss the scalings necessary when sampling lattice variables from Gaussian
distributions.  For completeness we list all possible types of initial conditions; however, in the body
of the text we use exclusively initial conditions based on current fluctuations.  We then give some
more details about the precise implementation of the current fluctuation initial conditions used 
in the body of the text.

It is common to use uncorrelated Gaussian random noise as the initial condition for either fields
or current fluctuations.  In the case of uncorrelated transverse vector potentials, for example, one assumes
that in the continuum limit
\bea
&& \langle A_i^a(\tau_0,{\bf x}_1,\eta_1) A_j^b(\tau_0,{\bf x}_2,\eta_2) \rangle =
\nonumber \\ && \hspace{2cm} 
\Delta^2 \delta^{a b} \delta_{ij} \delta^{(2)}({\bf x}_1 - {\bf x}_2) \delta(\eta_1 -\eta_2) \,,
\eea
where ${\bf x^\perp} \equiv (x,y)$ is a purely transverse two-vector.
In order to translate this statement into something useful for the lattice initial conditions we should convert to dimensionless variables
on the left and right hand sides.  In doing so we make use of the rescalings
specified in Eqs.~(\ref{eq:spatial-rescalings}) and (\ref{eq:field-rescalings})
and the Dirac delta function identity $\delta(ax) = \delta(x)/|a|$
to obtain (in terms of the lattice variables introduced in Eqs.~(\ref{eq:field-rescalings}) and (\ref{eq:current-rescalings}))
\begin{align}
\langle A_i^a(\tau_0,{\bf x}_1^\perp,\eta_1) A_j^b(\tau_0,{\bf x}_2^\perp,\eta_2) \rangle = 
\frac{g^2\Delta^2}{a_\eta} \delta^{a b} \delta_{ij} \delta_{x_1^\perp x_2^\perp} 
\delta_{\eta_1 \eta_2} \, .
\end{align}
In practice, this means that the $A_i$ variables should be Gaussian random numbers with a standard deviation
of $\sigma = g \Delta/a_\eta^{1/2}$.

\begin{table}[t]
\centering
\begin{tabular}{c c}
\hline
Case & Std. Dev. ($\sigma$) \\ [1ex] 
\hline       
Transverse Vector Potential ($A_i$) & $g \Delta/a_\eta^{1/2}$ \\ 
Longitudinal Vector Potential ($A_\eta$) & $g \Delta/(a a_\eta^{1/2})$ \\ [1ex] 
Transverse Conjugate Momentum ($\Pi_i$) & $g \Delta/a_\eta^{1/2}$ \\ [1ex] 
Longitudinal Conjugate Momentum ($\Pi_\eta$) & $g a\Delta/a_\eta^{1/2}$ \\ [1ex] 
Current fluctuations (${\cal W}$) & $\Delta/a_\eta^{1/2}$ \\ [1ex]
\hline
\end{tabular}
\caption[IC standard deviation]{Transverse and longitudinal lattice spacing scaling for a variety
of different initial condition types.}
\label{tab:stdev}
\end{table}

Using similar arguments we can derive the following lattice correlation functions in the case that we initialize 
longitudinal vector potentials
\begin{align}
\langle A_\eta^a(\tau_0,{\bf x}_1^\perp,\eta_1) A_\eta^b(\tau_0,{\bf x}_2^\perp,\eta_2) \rangle = 
\frac{g^2\Delta^2}{a^2 a_\eta} \delta^{a b} \delta_{x_1^\perp x_2^\perp} 
\delta_{\eta_1 \eta_2} \, ,
\end{align}
or transverse momenta
\begin{align}
\langle \Pi_i^a(\tau_0,{\bf x}_1^\perp,\eta_1) \Pi_j^b(\tau_0,{\bf x}_2^\perp,\eta_2) \rangle = 
\frac{g^2\Delta^2}{a_\eta} \delta^{a b} \delta_{ij} \delta_{x_1^\perp x_2^\perp} 
\delta_{\eta_1 \eta_2}  \, ,
\end{align}
or longitudinal momenta
\begin{align}
\langle \Pi_\eta^a(\tau_0,{\bf x}_1^\perp,\eta_1) \Pi_\eta^b(\tau_0,{\bf x}_2^\perp,\eta_2) \rangle = 
\frac{g^2a^2\Delta^2}{a_\eta} \delta^{a b} \delta_{x_1x_2} 
\delta_{\eta_1 \eta_2} \, , 
\end{align}
or auxiliary fields
\bea
&& \langle W^{a\alpha}(\tau_0,{\bf x}_1^\perp,\eta_1;\phi_1,y_1)
 W^{b\beta}(\tau_0,{\bf x}_2^\perp,\eta_2;\phi_2, y_2) \rangle = 
\nonumber \\ && 
\frac{\Delta^2}{a_\eta} \delta^{a b} \delta^{\alpha \beta} \delta_{x_1^\perp x_2^\perp} 
\delta_{y_1 y_2} \delta_{\eta_1 \eta_2}
\delta_{\phi_1 \phi_2}
 \, .
\eea
To summarize, when using Gaussian random initial conditions on anisotropic lattices, one should choose 
the standard deviations shown in Table \ref{tab:stdev}.
Moreover, unless initial fluctuations are only set up for the
gauge fields $A_i^a$ and $A_\eta^a$, a projection to satisfy
the Gauss law constraint (\ref{gausslawcontinuum3d}) 
is needed. In our simulations we
have however used a different setup which we now discuss.

\subsection{Initial Condition Setup}

The analytic study of
collective modes in anisotropically expanding ultarelativistic plasmas
\cite{Rebhan:2009ku} has found that the initial fluctuations in (only) induced
currents versus only initial fluctuations in collective fields reduces
considerably the delay of the onset of the plasma instabilities. 
As discussed in Sect.\ \ref{sec:initialconditions-continuum},
this means that such initial conditions dominate over
all other possibilities, and it is therefore sufficient to
concentrate on initial fluctuations in the $W$ fields which
directly encode the induced currents.




\begin{figure*}[t!]
\subfigure[~Variation of the initial random seed used for the current fluctuations.
All parameters are the same as in Fig.~\ref{fig:eDensity1}.]{
\includegraphics[width=0.44\textwidth]{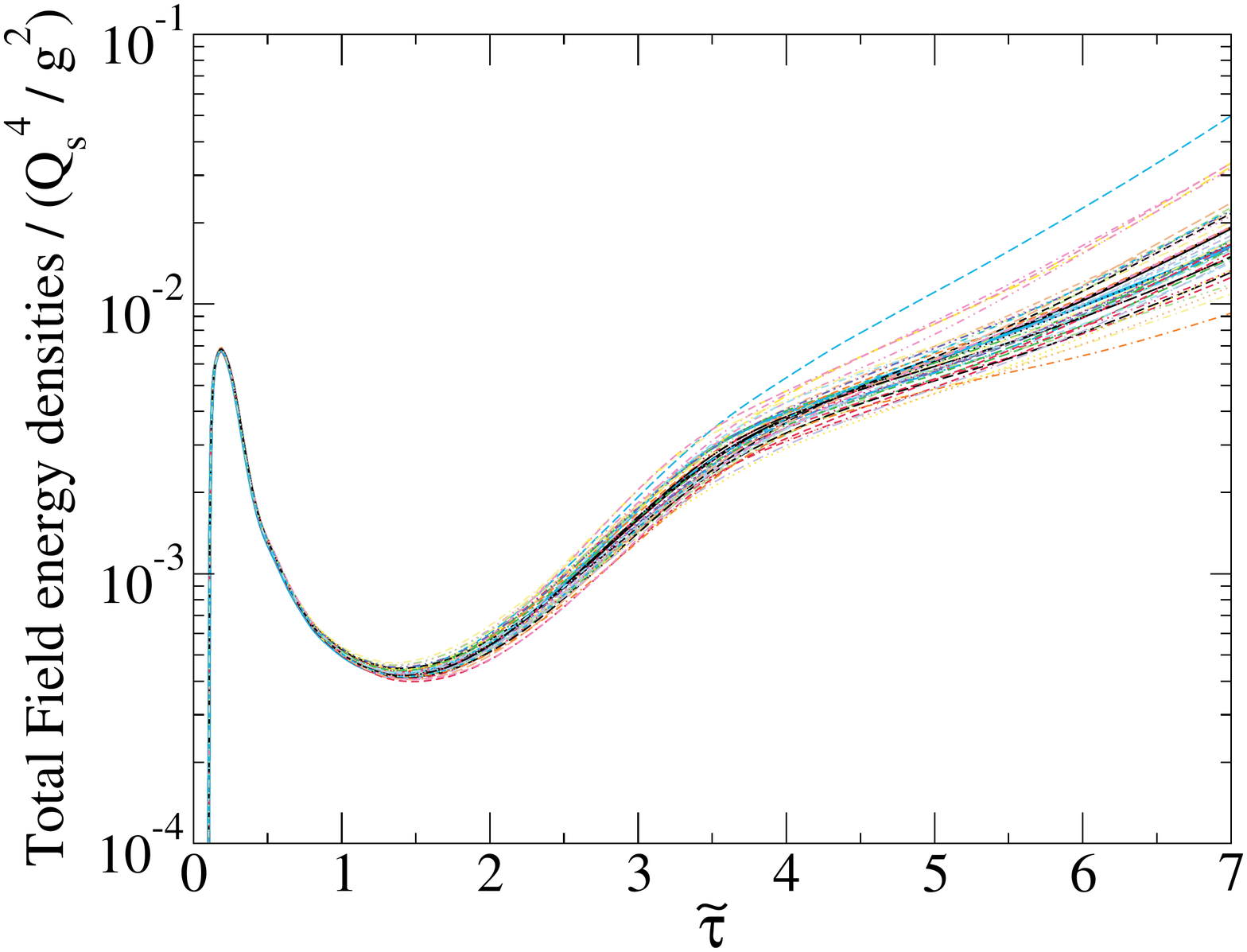}
\label{fig:numtests-collection-a}
} \hspace{5mm}
\subfigure[~Variation of the longitudinal cutoff, $\Lambda_\nu \nu_{\rm min}$ with $\nu_{\rm min}=1.96$, for the current fluctuation
initial conditions. All parameters except $\Lambda_\nu$ are the same as in Fig.~\ref{fig:eDensity1}.]{
\includegraphics[width=0.44\textwidth]{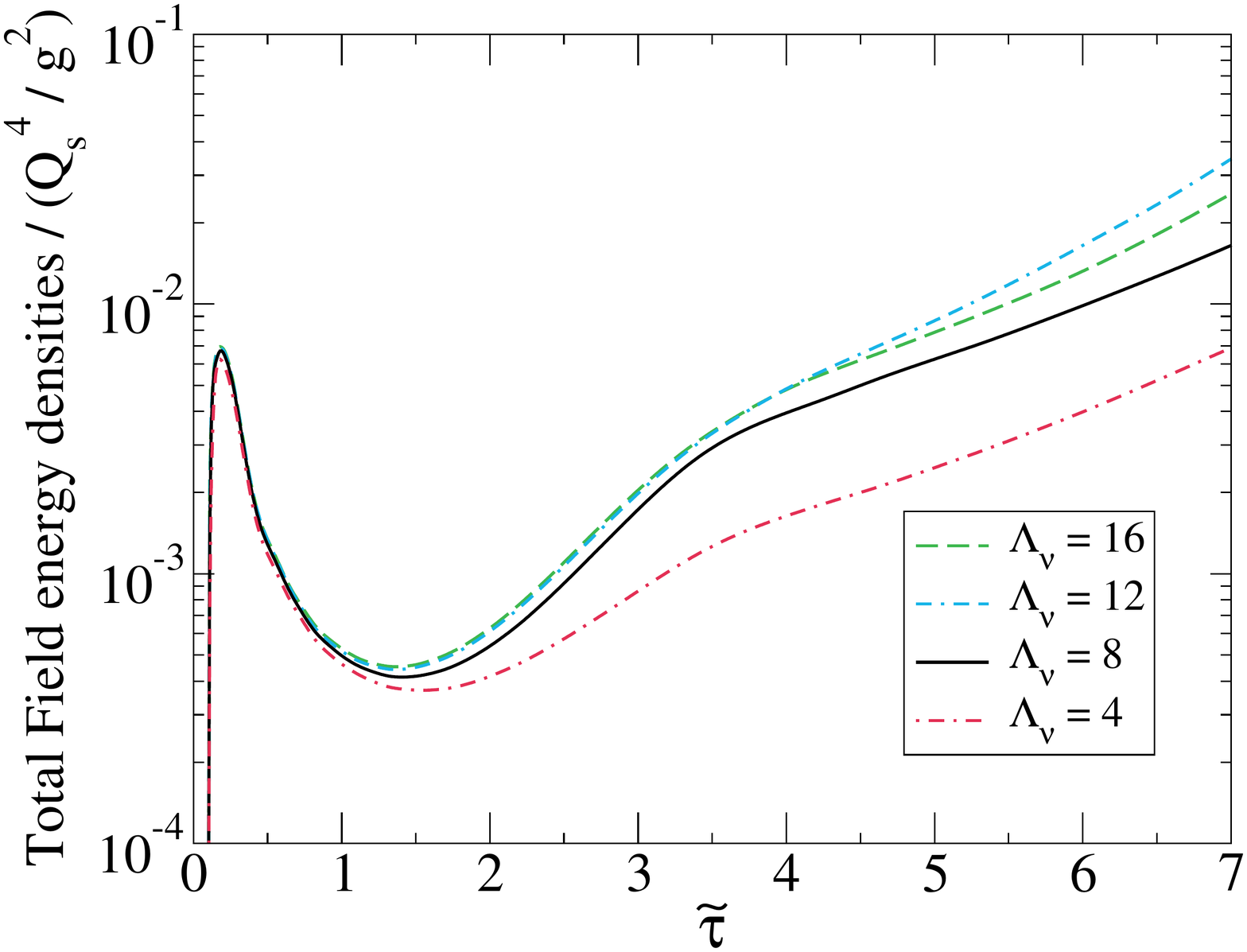}
\label{fig:numtests-collection-b}
}\\
\subfigure[~Variation of the transverse lattice spacing, $a$, while keeping the transverse 
lattice size, $L_T = N_T \, a$, fixed.  All parameters except $a$ and $N_T$ are the same as in Fig.~\ref{fig:eDensity1}.]{
\includegraphics[width=0.44\textwidth]{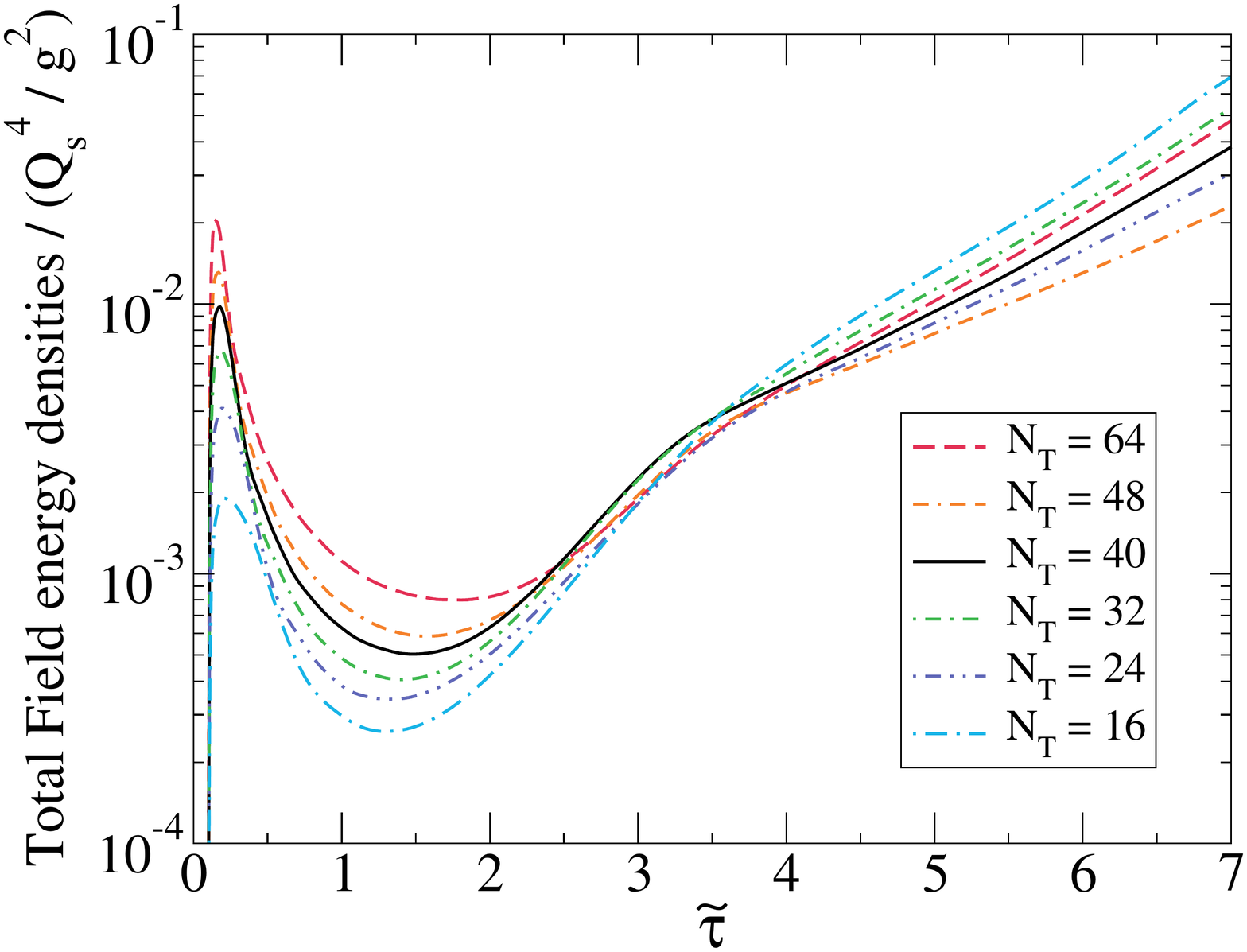}
\label{fig:numtests-collection-c}
} \hspace{5mm}
\subfigure[~Variation of the transverse lattice size, $L_T = N_T \, a$, while keeping the 
transverse lattice spacing, $a$, fixed.  All parameters except $N_T$ are the same as in Fig.~\ref{fig:eDensity1}.]{
\includegraphics[width=0.44\textwidth]{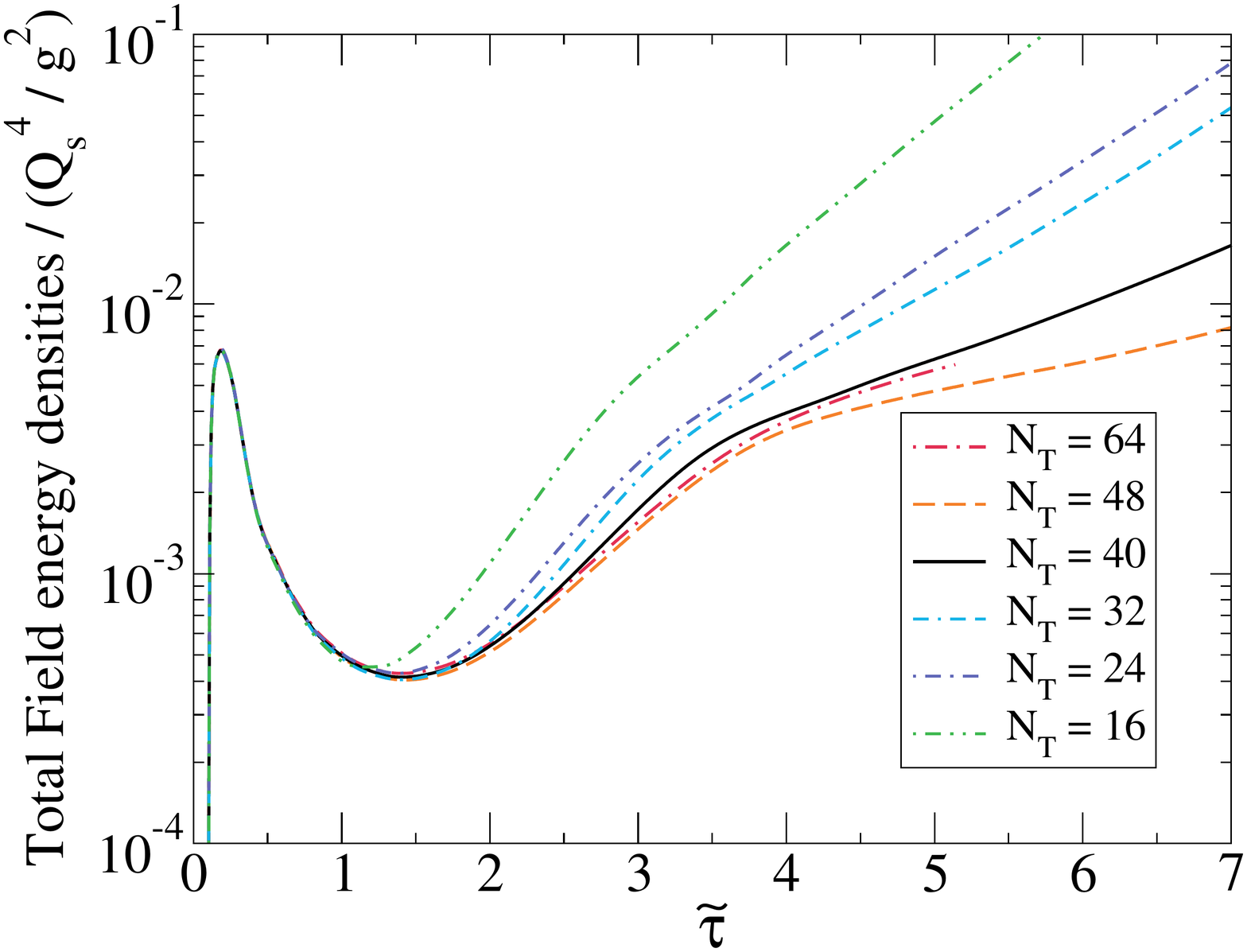}
\label{fig:numtests-collection-d}
}
\subfigure[~Variation of the longitudinal lattice spacing, $a_\eta$, while keeping the 
longitudinal lattice size, $L_\eta = N_\eta \, a_\eta$, fixed.  All parameters except $a_\eta$ and $N_\eta$ are the same
as in Fig.~\ref{fig:eDensity1}.]{
\includegraphics[width=0.44\textwidth]{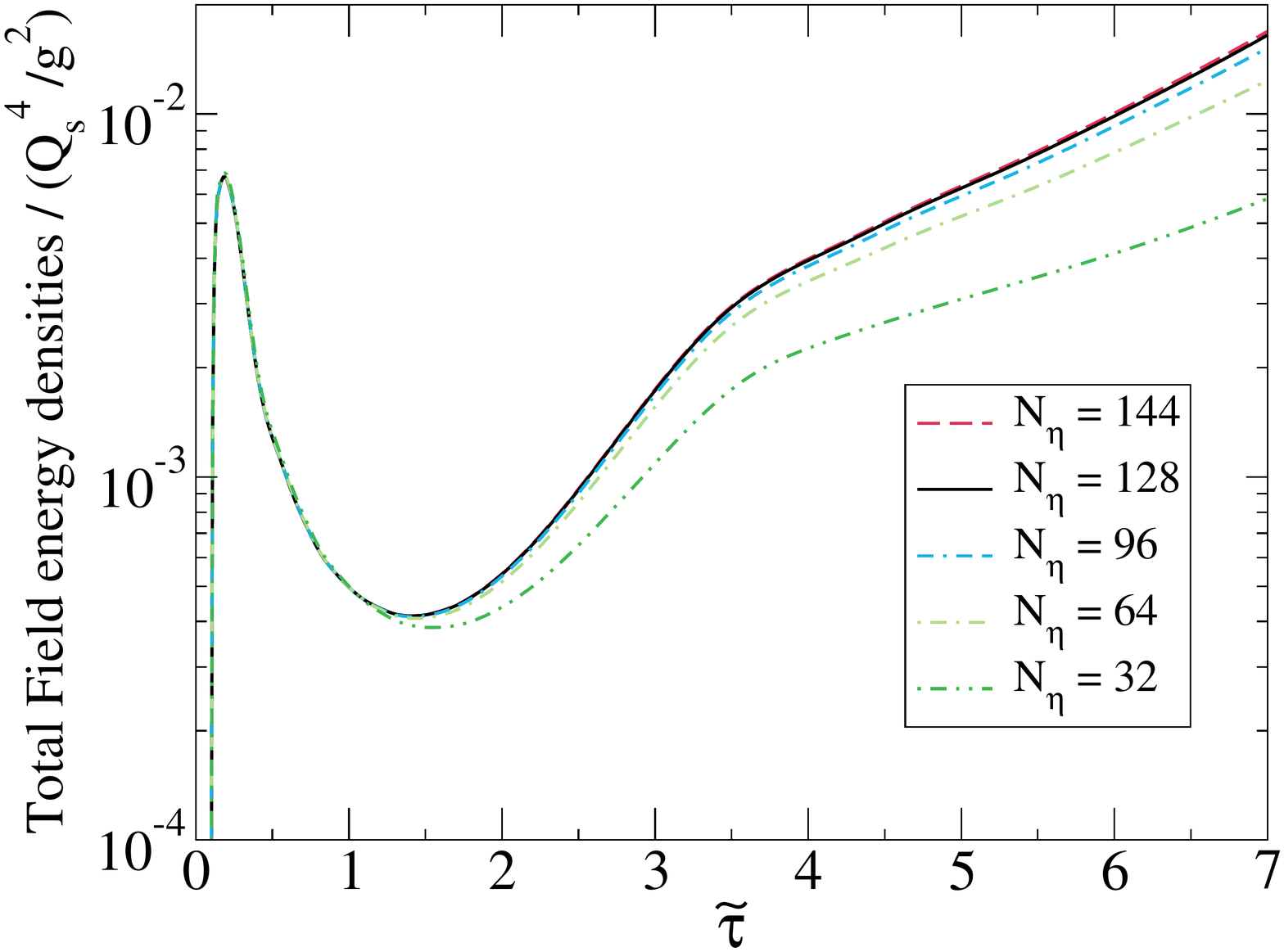}
\label{fig:numtests-collection-e}
} \hspace{5mm}
\subfigure[~Variation of the longitudinal lattice size, $L_\eta = N_\eta \, a_\eta$, while keeping 
the longitudinal lattice spacing, $a_\eta$, fixed.  All parameters except $N_\eta$ and $\Delta$ are the same as in 
Fig.~\ref{fig:eDensity1}.]{
\includegraphics[width=0.44\textwidth]{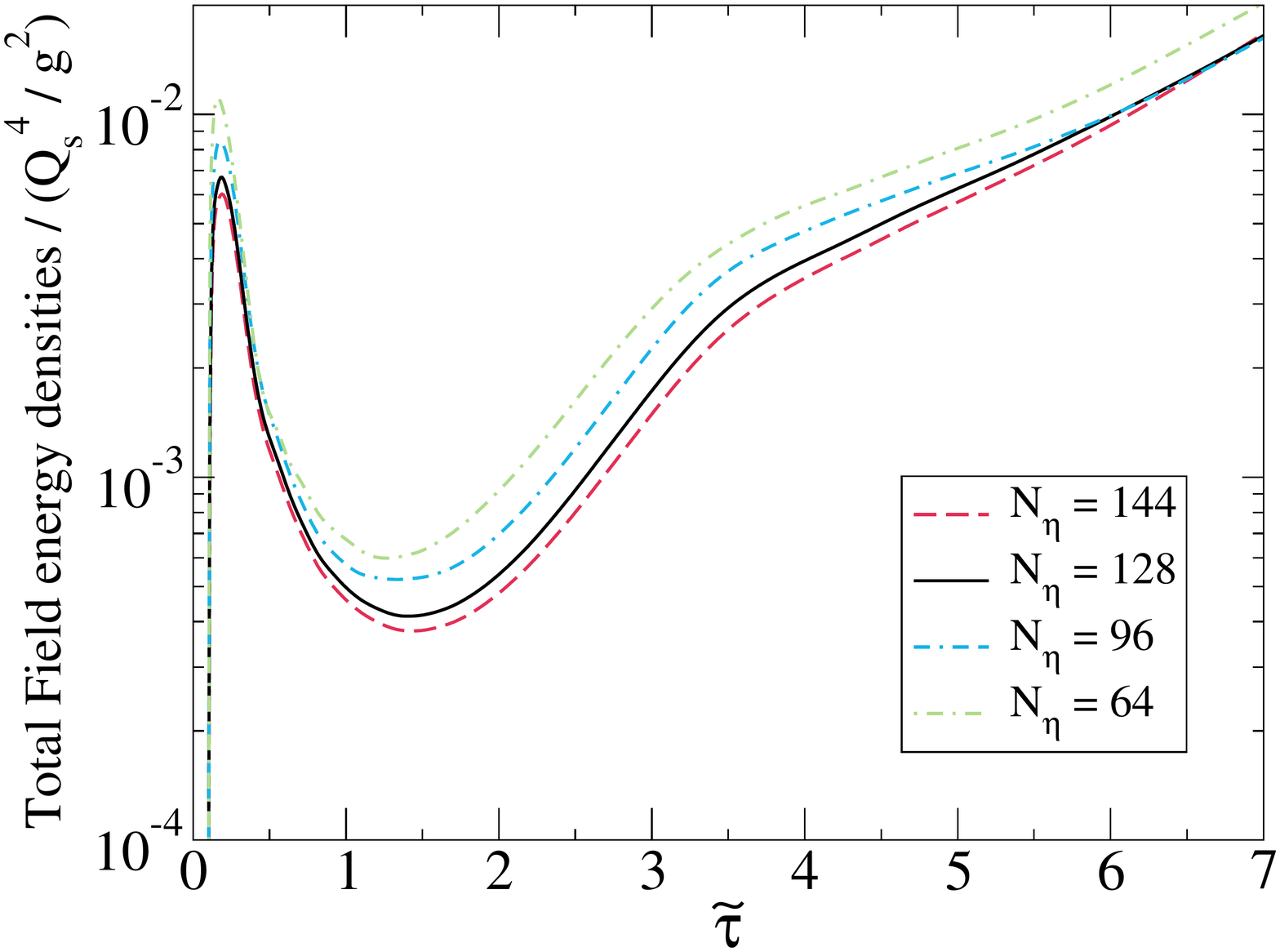}
\label{fig:numtests-collection-f}
}
\caption{(Color online) Collected numerical tests of unstable mode growth.   Each subpanel shows the chromofield
total energy density evolution subject to variation of various parameters.  The subcaptions contain a description of the 
parameters which are varied.}
\label{fig:numtests-collection}
\end{figure*}

\subsubsection{Longitudinal Current Initial Conditions}

For oblate anisotropy,
fluctuations in longitudinal currents give rise to stable plasmon modes,
and in the Abelian case they do not lead to any plasma instabilities.
We have used this to test our code for
unphysical instabilities (see App.\ \ref{app:numerics}).

The simplest initial fluctuations consistent with Gauss' law
which achieve this are fluctuations in only the $W^{a\eta}$ components
that are independent of $\phi$ and $y$ (thereby ensuring that
initially $j^\tau=0$) 
but nothing else
\begin{align} \label{initcondws}
&& \langle W^{a\eta}(\tau_0,{\bf x}_1^\perp,\eta_1;\phi_1,y_1)
 W^{b\eta}(\tau_0,{\bf x}_2^\perp,\eta_2;\phi_2, y_2) \rangle = 
\nonumber \\ 
 &&\frac{\Delta^2}{a_\eta} \delta^{a b}
\delta_{x_1^\perp x_2^\perp} 
\delta_{\eta_1 \eta_2}
\nonumber , \\
&&W^{ai} = 0, \quad 
U_{s+\frac{1}{2}} = {\mathbf 1}_{N_c} , 
\quad \Pi_{i,s} = \Pi_{\eta,s}=0.\quad 
\end{align}

\subsubsection{Transversal Current Initial Conditions}

In order to provide seed fields for Weibel instabilities,
longitudinal current fluctuations do not play an important
role (for oblate anisotropies). For simplicity we have therefore
only considered transverse current fluctuations by
only initializing $W^{ai}$ fields.
Because we have used rather fine lattices in the $\eta$ direction,
Gaussian random noise would correspond to very high UV
noise in longitudinal wave numbers even beyond the
scale which separates soft and hard modes, while hard modes
are already integrated out. We have therefore introduced a mode
number cutoff $\Lambda_\nu$ such that $\nu_{\rm max}=\Lambda_\nu \nu_{\rm min}$
with $\nu_{\rm min}= {2\pi}/({N_\eta a_\eta})$.
Again, the simplest initial fluctuations consistent with the Gauss law
are obtained by requiring that the $W^{ai}$ components
are independent of $\phi$ and $y$, and thus
initially $j^\tau=0$, while setting all other fields to zero initially.
This is now done in terms of the Fourier components 
$\tilde W^{ai}$ with
$W^{ai}(\ldots,\eta,\ldots)
=\sum_\nu \tilde W^{ai}(\ldots,\nu,\ldots)e^{i\nu\eta}$ according to

\begin{align} \label{initcondwu}
&& \langle \tilde W^{ai}(\tau_0,{\bf x}_1^\perp,\nu_1;\phi_1,y_1)
 \tilde W^{bj}(\tau_0,{\bf x}_2^\perp,\nu_2;\phi_2, y_2) \rangle = 
\nonumber \\ 
&&
\Delta^2
\delta^{a b} \delta^{ij} \delta_{x_1^\perp x_2^\perp} 
\delta_{\nu_1,-\nu_2}\theta(\nu_{\rm max}-|\nu_1|),
\nonumber \\
&&W^{a\eta} = 0, \quad U_{s+\frac{1}{2}} = {\mathbf 1}_{N_c}
, 
\quad \Pi_{i,s} = \Pi_{\eta,s}=0.\quad 
\end{align}

\section{Numerical Tests}
\label{app:numerics}

In this section we collect various numerical tests such as varying the lattice spacing, lattice size, spectral cutoffs, and
velocity-space resolution.  In Fig.~\ref{fig:numtests-collection} we collect six different tests.  
The variation with the random seed used for 
generating the necessary pseudorandom numbers used in the initial conditions is shown in 
Fig.\ref{fig:numtests-collection-a}.  As we can see from this figure there is a fair amount of variation with the random seed 
used; however, the results are all qualitatively the same.  In the results 
section our main results are averaged over the set of runs shown in 
Fig.\ref{fig:numtests-collection-a}.

The variation with the ultraviolet longitudinal mode cutoff used for initializing the initial current fluctuations via the 
auxiliary ${\cal W}$ fields is shown in Fig.\ref{fig:numtests-collection-b}.  As can be seen from this figure there is a
rapid convergence as the ultraviolet cutoff, $\Lambda_\nu \nu_{\rm min}$, is increased.  The set of runs 
shown in the main body of the text uses $\Lambda_\nu = 8$.

The variation with the transverse lattice spacing while holding the transverse lattice size fixed is shown in
Fig.\ref{fig:numtests-collection-c}.  This represents a test of the approach to the continuum as the transverse
lattice resolution is increased.  In the transverse plane we sample Gaussian random numbers which means as
the lattice spacing decreases the transverse configurations will be dominated by the high transverse momentum
part of the fluctuations.  This is evidenced by the fact that the initial energy density deposited in the fields by
the current fluctuations increases rapidly as one approaches the transverse continuum limit.  One could remove
this artifact by implementing a transverse mode cutoff on the lattice, but at this point in time we have not yet 
done so.  In the results section our standard set of runs uses $N_T = 40$.

The variation with the transverse lattice size while holding the transverse lattice spacing fixed is shown in
Fig.\ref{fig:numtests-collection-d}.  In this case we see a rather large effect.  In the limit that $N_T \rightarrow 1$
while holding $a$ fixed, one approaches a one-dimensional system which exhibits a faster growth rate due
to less mode competition.  We have verified that in this limit we reproduce our previously obtained results 
from Ref.~\cite{Rebhan:2008uj}.  The faster growth seen compared to Ref.~\cite{Rebhan:2008uj} is due
to the use of the more general initial conditions which include current fluctuations \cite{Rebhan:2009ku}.
In the results section our standard set of runs uses $N_T = 40$.

\begin{figure}[h]
\includegraphics[width=0.45\textwidth]{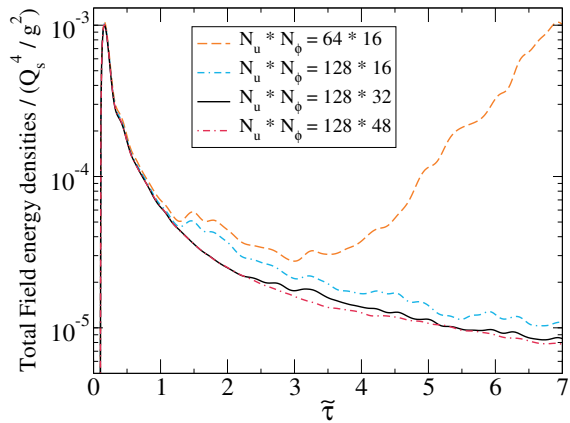}
\caption{(Color online) 
Evolution of a stable configuration initialized with Abelian longitudinal currents for different sized velocity lattices.
}
\label{fig:numtest-stable}
\end{figure}

The variation with the longitudinal lattice spacing while holding the longitudinal lattice size fixed is shown in
Fig.\ref{fig:numtests-collection-e}.  Due to the fact that we have implemented an ultraviolet cutoff on
fluctuations in the $\eta$-direction, we see a very nice convergence as the lattice resolution in the $\eta$
direction is increased.  In the results section our standard set of runs uses $N_T = 40$.
In the results section our standard set of runs uses $N_\eta = 128$.

The variation with the longitudinal lattice size while holding the longitudinal lattice spacing fixed is shown
in Fig.\ref{fig:numtests-collection-f}. (Here $\Delta$ has been adjusted to
correct for the different initial spectrum which starts at smaller $\nu_{\rm min}$ with larger $N_\eta$, leading to different initial energy densities.)
Once again we see only small variation with the assumed longitudinal
lattice size, with the late-time variations being consistent with those coming from random seed variation.
In the results section our standard set of runs uses $N_\eta = 128$.

Finally, in Fig.~\ref{fig:numtest-stable} we show the evolution of a stable Abelian configuration initialized
with Abelian longitudinal currents for various different velocity lattice resolutions $N_u \times N_\phi \in \{ 64 \times 16, 128 \times 16, 128 \times 32,
128 \times 48 \}$.  For this simulation the lattice spatial size was $N_T^2 
\times N_\eta = 32^2 \times 32$ with transverse lattice spacing of $a = 0.1$ fm and longitudinal lattice spacing of 
$a_\eta = $0.025.  The initial time was taken to be $\tau_0 = $ 0.1 fm/c and we used $\tau_{\rm iso}$ = 0.01 fm/c.  
For the temporal time step we use $\epsilon = 10^{-3}$ fm/c.  
With these initial conditions, the field energy should decay steadily after
the initial peak. 
This test turns out to be very sensitive
to the velocity-space resolution, i.e.\ the number of $\mathcal W$ fields.  
If this resolution is too crude, the field energy even grows at late times.
Fig.~\ref{fig:numtest-stable} shows that with a velocity lattice size of $N_u \times N_\phi = 
128 \times 32$ there is already good convergence to the correct time evolution of the system.  Unstable modes are in fact less sensitive to the 
velocity resolution in the $\phi$ direction; however, being cautious we have performed all simulations using $N_u \times 
N_\phi = 128 \times 32$.

\bibliography{hel3d}

\end{document}